\documentclass[oneolumn]{aastex631}

\newcommand{\kmsmpc}{\rm km~s^{-1}~Mpc^{-1}}
\newcommand{\dn}{D_{n}4000}

\usepackage{color}
\usepackage{multirow}
\usepackage{amsmath}
\usepackage{appendix}
\usepackage{hhline}
\usepackage{hyperref}

\shorttitle{HectoMAP summed spectra}
\shortauthors{Damjanov et al.}

\graphicspath{{./}{figures/}}

\begin{document}

\title{The average stellar population age and metallicity of intermediate-redshift quiescent galaxies }

\author{Ivana Damjanov}
\affil{Department of Astronomy and Physics, Saint Mary's University, 923 Robie Street, Halifax, NS B3H 3C3, Canada; \href{mailto:Ivana.Damjanov@smu.ca}{Ivana.Damjanov@smu.ca}}
\affil{Canada Research Chair in Astronomy and Astrophysics, Tier II}
\affil{Brain Pool Fellow, National Research Foundation of Korea}
\author{Margaret J. Geller}
\affil{Center for Astrophysics | Harvard \& Smithsonian, 60 Garden Street, Cambridge, MA 02138, USA}
\author{Jubee Sohn}
\affil{Astronomy Program, Department of Physics and Astronomy, Seoul National University, 1 Gwanak-ro, Gwanak-gu, Seoul 08826, Republic of Korea} 
\affil{SNU Astronomy Research Center,
Seoul National University, Seoul 08826, Republic of Korea}

\begin{abstract}
The HectoMAP spectroscopic survey provides a unique mass-limited sample of more than 35,000 quiescent galaxies ($\dn>1.5$) covering the redshift range $0.2<z<0.6$. We segregate galaxies in bins of properties based on stellar mass, $\dn$, and redshift to construct a set of high signal-to-noise spectra representing massive ($M_\ast>10^{10}\,M_\sun$) quiescent population at intermediate redshift. These high-quality summed spectra enable full spectrum fitting and the related extraction of the average stellar population age and metallicity. The average galaxy age increases with the central $\dn$ as expected. The correlation is essentially invariant with stellar mass; thus $\dn$ is a robust proxy for quiescent galaxy stellar population age. HectoMAP provides the first quiescent sample at intermediate redshift comparable with $z\sim0$ mass-complete datasets. Scaling relations derived from the HectoMAP summed spectra connect stellar age and metallicity with quiescent galaxy stellar mass up to $z\sim0.5$. Anti-correlation between the equivalent width of the [O II] emission line and stellar age, together with the mild increase in stellar age with stellar mass, supports a broad range of timescales for the mass assembly of intermediate-redshift quiescent systems. On average, the most massive galaxies ($M_\ast>10^{11}\, M_\sun$)  assemble the bulk of their stars at earlier epochs. A strong increase in the average stellar metallicity with stellar mass, along with the correlation between the [O II] equivalent width and metallicity at $0.2<z<0.4$, suggests that lower-mass galaxies are more likely to have recent star formation episodes; related feedback from massive stars affects the chemical enrichment of these galaxies. 

\end{abstract}


\section{Introduction} \label{sec:intro}

Processes that drive the growth of galaxy stellar mass, including the accretion, in-situ consumption and removal of gas, and the addition of ex-situ formed stars through interactions, determine the distribution of observed galaxy stellar population properties. The fractional contribution of stellar populations with various chemical compositions and ages to the stellar mass of quiescent galaxies holds information about the suite of processes that contribute to galaxy mass assembly throughout cosmic history. Measurements of average quiescent galaxy age and metallicity and their dependence on other fundamental galaxy properties over a broad redshift range constrain the main drivers of galaxy mass assembly \citep[e.g.,][]{Trager2000, Schiavon2006, Kaviraj2009, Gallazzi2014, Pipino2014, Scott2017, Wu2018, Li2018, Chauke2019, Neumann2021, Lu2023, Cappellari2023}. 

Different stellar population properties have similar effects on the galaxy spectral energy distribution (SED). Contributions from either older or more metal-rich stellar population redden the integrated SED \citep[age-metallicity degeneracy,][]{Worthey1994}. In contrast, galaxy spectra in rest-frame visible light are rich in features with different sensitivity to stellar population age and chemical enrichment. At moderate signal-to-noise ratio and spectral resolution (SNR$\gtrsim20$ and $R\gtrsim1000$, respectively), full spectrum fitting in rest-frame visible light enables robust separation of age and metallicity \citep[][and references therein]{Conroy2013}. 

Dense spectroscopic surveys, critical tools for mapping large scale structure, provide statistical samples for studies of galaxy stellar population properties \citep{Geller2015}. The Sloan Digital Sky Survey, the largest spectroscopic survey of the low-redshift ($z<0.2$) universe \citep[SDSS;][]{York2000}, and its mass-complete galaxy samples are the benchmark for scaling relations between galaxy age and metallicity as a function of stellar mass and/or velocity dispersion \citep[e.g.,][]{Gallazzi2005, CidFernandes2005, Panter2008, Thomas2010, Conroy2014, Gallazzi2014, Peng2015, Citro2016, Zahid2017, Trussler2020, Sextl2023}. 

The increase in average galaxy age with stellar mass at $z\sim0$ supports galaxy mass ``downsizing in time";  more massive galaxies form the bulk of their stellar mass earlier in cosmic history \citep{Cowie1996, Fontanot2009}. The increase of stellar metallicity with stellar mass in local star-forming galaxies suggests that the stellar-to-gas mass ratio is the main driver of chemical evolution \citep{Zahid2017, Wu2017}. Low-redshift quiescent galaxies follow a similar relation with the mass-dependent offset towards higher metallicities \citep{Peng2015}. Quenching mechanisms that include both the suppression of gas accretion from the circumgalactic/intergalactic medium and outflows from galaxies with lower stellar masses reproduce the metallicities of $z\sim0$ quiescent galaxies \citep[e.g.,][]{Larson1974, Tremonti2004, Trussler2020}.   

At $z<1$ quiescent galaxies dominate the integrated stellar mass density \citep[e.g.,][]{McLeod2021}. Observed trends in stellar population properties of quiescent population in this redshift range trace the modes and timescales of stellar mass assembly over a significant portion of cosmic history. However, the quality of spectra in large-area surveys that reach intermediate redshift ($0.2<z\lesssim0.7$) is not sufficient to support full spectrum fitting applied to  {\it individual} galaxies \citep{Costantin2019}. 

Spectra with SNR~$\lesssim10$ and $R\sim1000$ in large spectroscopic surveys of the low and intermediate-redshift universe \citep[e.g.,][]{Strauss2002, Colless2003, Ahn2014, Alam2015, Liske2015, Geller2016} enable measurements of the $\dn$ spectral index, a metallicity-dependent proxy for the stellar population age \citep[e.g.,][]{Kauffmann2003}. \citet{Damjanov2023} use $\dn$ for $0.2<z<0.6$ quiescent galaxies in the HectoMAP survey \citep{Sohn2021, Sohn2023} as the evolutionary link between quiescent galaxies at high-redshift end of the survey ($z\sim0.6$) and their descendants down to $z\sim0.2$. Full spectrum fitting of a set of high-quality spectra can further explore $\dn$ as a tracer of the relative change in average quiescent stellar population age.

Stacking spectra enables full spectrum fitting of representative quiescent galaxy spectra at intermediate redshift \citep[e.g.,][]{Choi2014}. The resulting fractional contribution of simple stellar populations (SSP) with different ages and chemical compositions to the average spectra of galaxies segregated according to spectro-photometric properties provides estimates of the average stellar population age and metallicity. These estimates test age proxies measured from individual spectra (e.g., $\dn$). Scaling relations between galaxy stellar population properties based on spectral fitting and other spectral indicators at redshift $z>0.2$ constrain timescales of the processes that define those relations. 

Homogeneous samples that cover a broad redshift range are critical for studying evolution in galaxy stellar population properties. The absolute values of galaxy stellar ages and metallicities depend sensitively on the models, methodology, and spectral range \citep{Saracco2023}. Stellar mass-limited samples are necessary for constraining the evolution in the properties of galaxy populations \citep[e.g.,][]{Damjanov2022}. Tracing {\it relative} variations in galaxy stellar population properties and their relations with e.g., stellar mass over cosmic time requires consistently derived estimates of average age and metallicity based on complete galaxy samples.

For a homogeneous mass-complete sample of $\sim36000$ quiescent ($\dn>1.5$) HectoMAP galaxies with stellar mass $M_\ast>10^{10}\, M_\sun$, we construct high SNR summed spectra of galaxy samples segregated by stellar mass and $\dn$ in $\Delta z=0.1$ bins covering  $0.2<z<0.6$ redshift. From full spectrum fitting to the stacked spectra, we derive the average stellar age and metallicity and explore their relations with other spectral indices measured directly from the stacked spectra. These stellar population properties enable a first direct comparison between the HectoMAP intermediate redshift scaling relations and similar results based on mass-complete SDSS quiescent samples \citep[e.g.,][]{Gallazzi2014, Peng2015, Citro2016, Trussler2020}. 

Section~\ref{data} describes the HectoMAP dataset. Section~\ref{proc} describes the detailed procedure for constructing summed spectra. Section~\ref{method} describes the full spectrum fitting approach we use to derive average stellar population properties from the summed spectra. Section~\ref{ages:features} investigates the correlations between stellar population ages and spectral indices ($\dn$, equivalent widths of [O II] and H$_{\delta_A}$). We construct and discuss the stellar age - stellar mass and the stellar metallicity - stellar mass scaling relations based on the mass-complete quiescent sample at $0.2<z<0.6$ in Section~\ref{evol:am}. In Section~\ref{discussion} we relate observed scaling relations to the timescales for mass assembly of quiescent galaxies at intermediate redshift. We conclude in Section~\ref{conclusions}. We adopt the Planck cosmological parameters \citep{Planck2016} with $H_{0} = 67.74~\kmsmpc$, $\Omega_{m} = 0.3089$, $\Omega_{\Lambda} = 0.6911$ and the AB magnitude system.       

\section{Summed Spectra}
\subsection{The HectoMAP Redshift Survey}\label{data}
HectoMAP is a panoramic, dense, red-selected redshift survey carried out with the Hectospec 300-fiber spectrograph on the 6.5 m MMT \citep{Fabricant2005}. The 270 mm$^{-1}$ Hectospec grating yields a typical resolution of $\sim$6~\AA\ and covers the wavelength range 3700–9100~\AA. We sum these spectra in bins of redshift, $\dn$, and Stellar mass to trace the spectral evolution of the quiescent galaxy population over the redshift range 0.2$ < z < 0.6$

HectoMAP covers a total area of 55 deg$^{2}$ in a 1.5 deg wide strip at high declination within the limits: $200 < \rm{R.A. (deg)} < 250$ and $42.5 < \rm{decl. (deg)} < 44.0$ \citep{Sohn2021,Sohn2023}. The survey includes 95,403 distinct galaxies with a redshift. Table 1 of \citet{Sohn2023} lists the parameters that define the survey.

The ultimate photometric calibration of HectoMAP completeness is based on Sloan Digital Sky Survey (SDSS) DR16 \citep{Ahumada2020}. HectoMAP includes a bright survey with $r_{petro,0} < 20.5$ and 
$(g - r)_{model,0} > 1$. The faint portion of the survey has $20.5 < r_{petro,0} \leq 21.3$ and red-selection $(g - r)_{model,0} > 1$ and $(r - i)_{model,0} > 0.5$ where $r_{petro,0}$ is the SDSS Petrosian magnitude corrected for Galactic extinction. Color selection is based on the extinction-corrected SDSS model magnitudes. For the faint portion, the $(r - i)_{model,0}$ cut removes low-redshift objects. We impose an additional surface brightness limit, $r_{fiber,0} < 22.0$, throughout the survey because red objects below this limit yield very low signal-to-noise spectra. The limiting $r_{fiber,0}$ magnitude corresponds to the extinction-corrected flux within the Hectospec fiber aperture.

\citet{Sohn2021} and \citet{Sohn2023} describe the two data releases that cover the entire HectoMAP redshift survey. The releases include the redshifts, $\dn$\footnote{$\dn$ is a ratio between fluxes (in $f_\nu$ units) measured in two narrow wavelength intervals: $4000-4100$~\AA\ and $3850-3950$~\AA\ \citep{Balogh1999}. It is lower for galaxies with spectra dominated by younger stellar populations \citep{Kauffmann2003}.}, and stellar masses along with the match to SDSS photometry. In addition, \citet{Damjanov2023} use Subaru HSC observations of a portion of the HectoMAP region to derive galaxy sizes.

HectoMAP includes $\sim2000$ redshifts per square degree. The median depth of the survey is $z = 0.345$. Within the color selection, the survey is 81\% uniformly complete to the limit $r_{petro,0} = 20.5$. For galaxies with $20.5 < r_{petro,0} \leq 21.3$, the sample is 72\% complete and the completeness is somewhat lower near the survey edges. 

The typical error in a HectoMAP redshift is 38 km s$^{-1}$ There is a small offset of 39 km s$^{-1}$ between Hectospec redshifts and the SDSS \citep{Sohn2021,Sohn2023}. This offset is irrelevant for the analysis here.

We compute stellar masses of HectoMAP galaxies based on SDSS DR16 $ugriz$ photometry. We use the Le Phare software package \citep{Arnouts1999,Ilbert2006} that incorporates stellar population synthesis models of \citet{Bruzual03}, a \citet{Chabrier03} IMF, and a \citet{Calzetti00} extinction law. We explore two metallicities (0.4 and 1 solar) and a set of exponentially decreasing star formation rates. We derive the mass-to-light ratio from the best model and we then use this ratio to convert the observed luminosity into a stellar mass.

\citet{Damjanov2023} construct complete stellar mass limited subsamples of quiescent HectOMAP galaxies covering the redshift range $0.2 < z < 0.6$. We follow their procedure to construct the subsampled we used to construct the summed spectra. There are 35996 galaxies in the stellar mass complete sample. 

\subsection{Procedure for Summing the Spectra}\label{proc}

\begin{deluxetable}{c c c c c c}\label{tab1}
\tabletypesize{\tiny}
\tablecaption{Number of individual HectoMAP galaxies in each summed spectrum}
\tablewidth{0pt}
\tablehead{
\colhead{Stellar mass range} &\multirow{2}{*}{$\dn$ range} &\multicolumn{4}{c}{Redshift interval}\\
\cline{3-6}
\colhead{[$\log(M_\ast/1\, M_\sun)$]}& & \colhead{$0.2<z<0.3$} &\colhead{$0.3<z<0.4$} & \colhead{$0.4<z<0.5$} &\colhead{$0.5<z<0.6$}
}
\startdata
\multirow{6}{*}{10.0--10.2} & 1.5--1.6 & 227 &&&\\
&1.6--1.7 & 222 & \nodata & \nodata & \nodata\\
&1.7--1.8 & 181 & \nodata & \nodata & \nodata\\
&1.8--1.9 & 113 & \nodata & \nodata & \nodata\\
&1.9--2.0 &  52 & \nodata & \nodata & \nodata\\
&$>$2     &  74 & \nodata & \nodata & \nodata\\
\hline
\multirow{6}{*}{10.2--10.4} & 1.5--1.6  & 450 & 153 &&\\
&1.6--1.7 & 433 & 140 & \nodata & \nodata\\
&1.7--1.8 & 423 & 121 & \nodata & \nodata\\
&1.8--1.9 & 335 &  88 & \nodata & \nodata\\
&1.9--2.0 & 190 &  31 & \nodata & \nodata\\
&$>$2     & 211 &  44 & \nodata & \nodata\\
\hline
\multirow{6}{*}{10.4--10.6} & 1.5--1.6  & 536 & 538 &&\\
&1.6--1.7 & 559 & 516 & \nodata & \nodata\\
&1.7--1.8 & 602 & 458 & \nodata & \nodata\\
&1.8--1.9 & 606 & 353 & \nodata & \nodata\\
&1.9--2.0 & 402 & 208 & \nodata & \nodata\\
&$>$2     & 363 & 244 & \nodata & \nodata\\
\hline
\multirow{6}{*}{10.6--10.8} & 1.5--1.6  & 415 & 659 & 184&\\
&1.6--1.7 & 500 & 733 & 185 & \nodata\\
&1.7--1.8 & 641 & 833 & 138 & \nodata\\
&1.8--1.9 & 723 & 687 & 137 & \nodata\\
&1.9--2.0 & 565 & 358 &  82 & \nodata\\
&$>$2     & 450 & 393 & 125 & \nodata\\
\hline
\multirow{6}{*}{10.8--11} & 1.5--1.6  & 283 & 544 & 444& 53\\
&1.6--1.7 & 342 & 566 & 459 & 58\\
&1.7--1.8 & 527 & 753 & 483 & 43\\
&1.8--1.9 & 706 & 759 & 384 & 35\\
&1.9--2.0 & 580 & 458 & 248 & 16\\
&$>$2     & 414 & 393 & 341 & 25\\
\hline
\multirow{6}{*}{11--11.2} & 1.5--1.6  & 102 & 261 & 298& 270\\
&1.6--1.7 & 139 & 348 & 341 & 313\\
&1.7--1.8 & 265 & 498 & 440 & 280\\
&1.8--1.9 & 456 & 577 & 389 & 207\\
&1.9--2.0 & 466 & 370 & 252 & 131\\
&$>$2     & 315 & 309 & 328 & 147\\
\hline
\multirow{6}{*}{11.2--11.4} & 1.5--1.6  & 25 & 70 & 92& 202\\
&1.6--1.7 &  52 & 120 & 155 & 305\\
&1.7--1.8 & 103 & 201 & 210 & 310\\
&1.8--1.9 & 198 & 305 & 263 & 268\\
&1.9--2.0 & 249 & 232 & 200 & 167\\
&$>$2     & 178 & 171 & 196 & 157\\
\hline
\multirow{6}{*}{11.4--11.6} & 1.5--1.6  & 6 & 18 & 21& 98\\
&1.6--1.7 &  13 &  27 &  55 & 160\\
&1.7--1.8 &  26 &  53 &  86 & 191\\
&1.8--1.9 &  47 &  96 & 108 & 176\\
&1.9--2.0 &  90 &  87 &  92 & 107\\
&$>$2     &  88 &  90 &  90 & 118\\
\enddata
\end{deluxetable}

We construct high signal-to-noise (SNR) summed spectra as a basis for full-spectrum fitting. We sum spectra of quiescent HectoMAP galaxies with similar redshift, stellar mass, and $\dn$. Our approach is similar to the procedure of \citet{Zahid2017} and \citet{Andrews2013}. Here we provide an overview and showcase a subset of the resulting summed spectra. 

We first segregate the galaxy spectra in $\Delta z=0.1$ and $\Delta(\log(M_\ast/M_\sun)=0.2$~dex abins of redshift and stellar mass, respectively; we include all quiescent ($\dn>1.5$) galaxies above the stellar mass limit of the survey \citep{Damjanov2023}. Within each redshift and stellar mass bin, we sort quiescent galaxies into bins with $\Delta(\dn)=0.1$ for $\dn<2.5$; we include all galaxies with $\dn>2$ in the largest value $\dn$ bin. Table~\ref{tab1} lists the number of individual galaxy spectra included in each stack.

Within each redshift interval, we shift the spectra to the rest-frame based on the measured redshift \citep{sohn2023}. We linearly interpolate between rest-frame wavelength limits for each spectrum with a wavelength pixel resolution $\Delta\lambda=1$~\AA. After normalizing each spectrum to the mean flux in the 4400-4450~\AA\ range, we co-add the spectra for each stellar mass and $\dn$ bin based on the mean flux in each wavelength pixel. We weight all of the spectra equally in calculating the average flux. 
 
\begin{figure*}[!h]
\centering
\includegraphics[width=0.8\textheight]{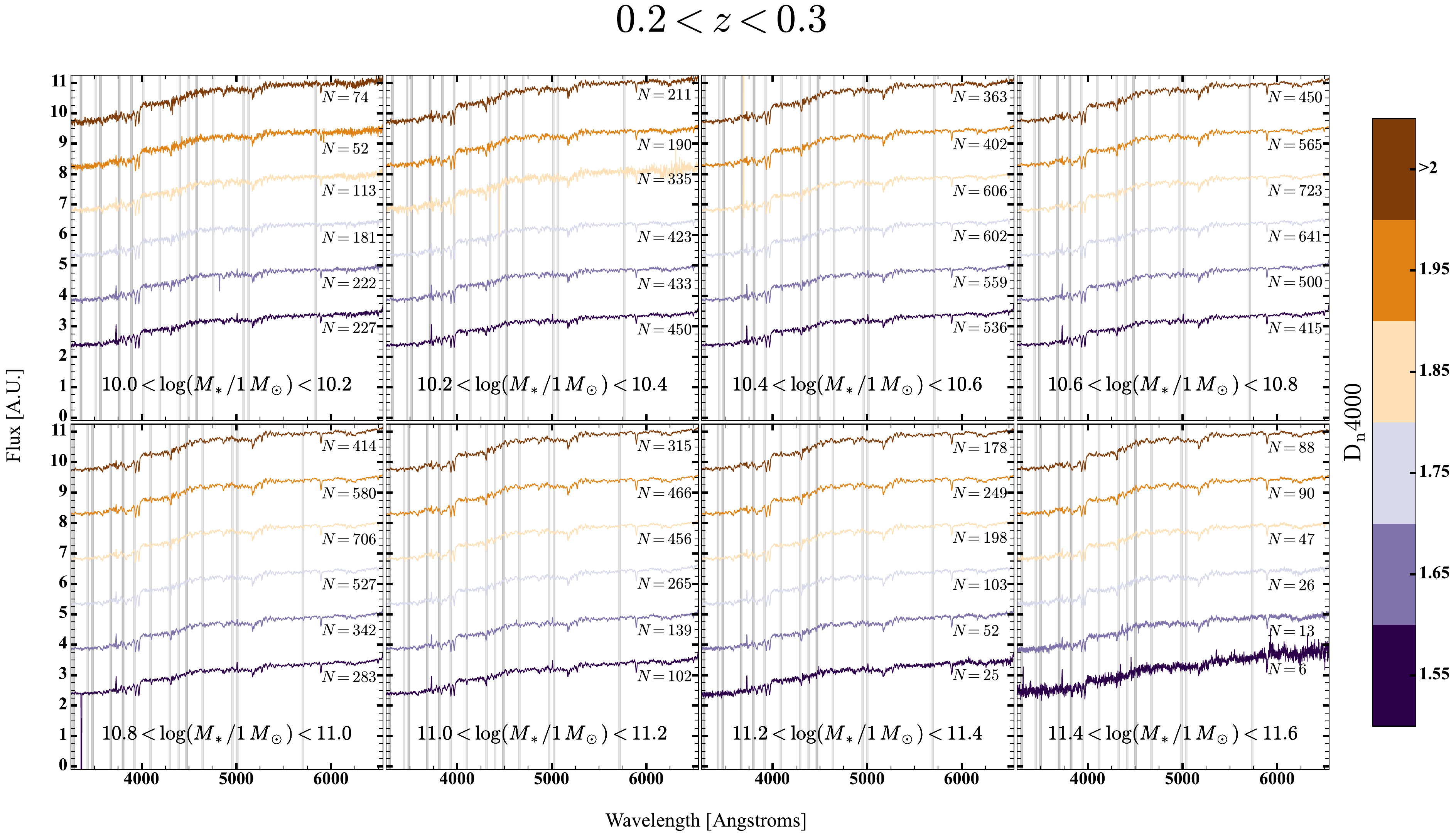}

\caption{Averaged summed spectra of HectoMAP quiescent galaxies with $0.2<z<0.3$ segregated by stellar mass (panels) and $\dn$ (colors). The number next to each spectrum indicates the total number of individual HectoMAP spectra in the stack. Gray lines show the positions of emission lines in the Mount Hopkins night sky spectrum \citep{Massey2000} in the rest frame of the $0.2<z<0.3$ stack. The average flux density of the co-adds is in arbitrary units.
\label{f1}}
\end{figure*}

Figure~\ref{f1} shows the summed spectra for the mass-limited sample of quiescent HectoMAP galaxies at $0.2<z<0.3$ segregated by stellar mass and $\dn$. We show summed spectra for the remaining three redshift bins of the HectoMAP survey ($0.3 < z < 0.4$, $0.42 < z < 0.5$\footnote{This redshift bin is slightly narrower because we omit the spectra in the redshift range $0.4<z<0.42$. For quiescent galaxies in this redshift range, the prominent absorption features, the Ca II K and H absorption lines ($\lambda= 3934~$\AA\ and $\lambda= 3969~$\AA, respectively), are significantly contaminated by the sky line [O I]$\lambda5577$.}, and $0.5<z<0.6$) in the Appendix~\ref{Spectra:other}.

We calculate the signal-to-noise ratio (SNR) Of the summed spectra in the rest-frame wavelength interval $3500-5500$ \AA\ using the \textsc{DER\_SNR} function\footnote{\url{https://specutils.readthedocs.io/en/stable/api/specutils.analysis.snr_derived.html}} of the \texttt{Specutils} Python software package. The SNR of the summed spectra is in the range $\sim10\lesssim\mathrm{SNR}\lesssim150$. The equivalent SNR for individual HectoMAP spectra is between 3 and 25; $\gtrsim75\%$ of the unsummed spectra have SNR$< 20$. Except for the $0.5<z<0.6$ redshift bin, where $\sim30\%$ of the summed spectra have $\mathrm{SNR} < 50$, most ($83\%$) of the summed spectra have $\mathrm{SNR}>50$. Between $M_\ast\sim2\times10^{10}\, M_\sun$ and $M_\ast\sim10^{11}\, M_\sun$, the average SNR of the summed spectra increases mildly with stellar mass (from SNR$\sim70$ to SNR$\sim90$). However, in all redshift intervals, the most massive galaxies ($M_\ast>2\times10^{11}\, M_\sun$) have the largest fraction of low-SNR spectra (SNR$<50$) and the lowest mean/median SNR ($\sim60$). The SNR of the summed spectra shows no trend with $\dn$. 

The Hectospec $0\farcs75$ fiber aperture covers different radii for galaxies at different redshift. We thus calculate the ratio between the projected fiber size and the median size of galaxies in each stellar mass and $\dn$ bin. We take the galaxy size as the half-light radius along the major axis ($R_{e}$) of the S\'ersic model fitted to the galaxy light profile in the Hyper Suprime-Cam (HSC) $i-$ band images \citep{Damjanov2023}. 

\begin{figure*}[!h]
\centering
\includegraphics[width=0.5\textheight]{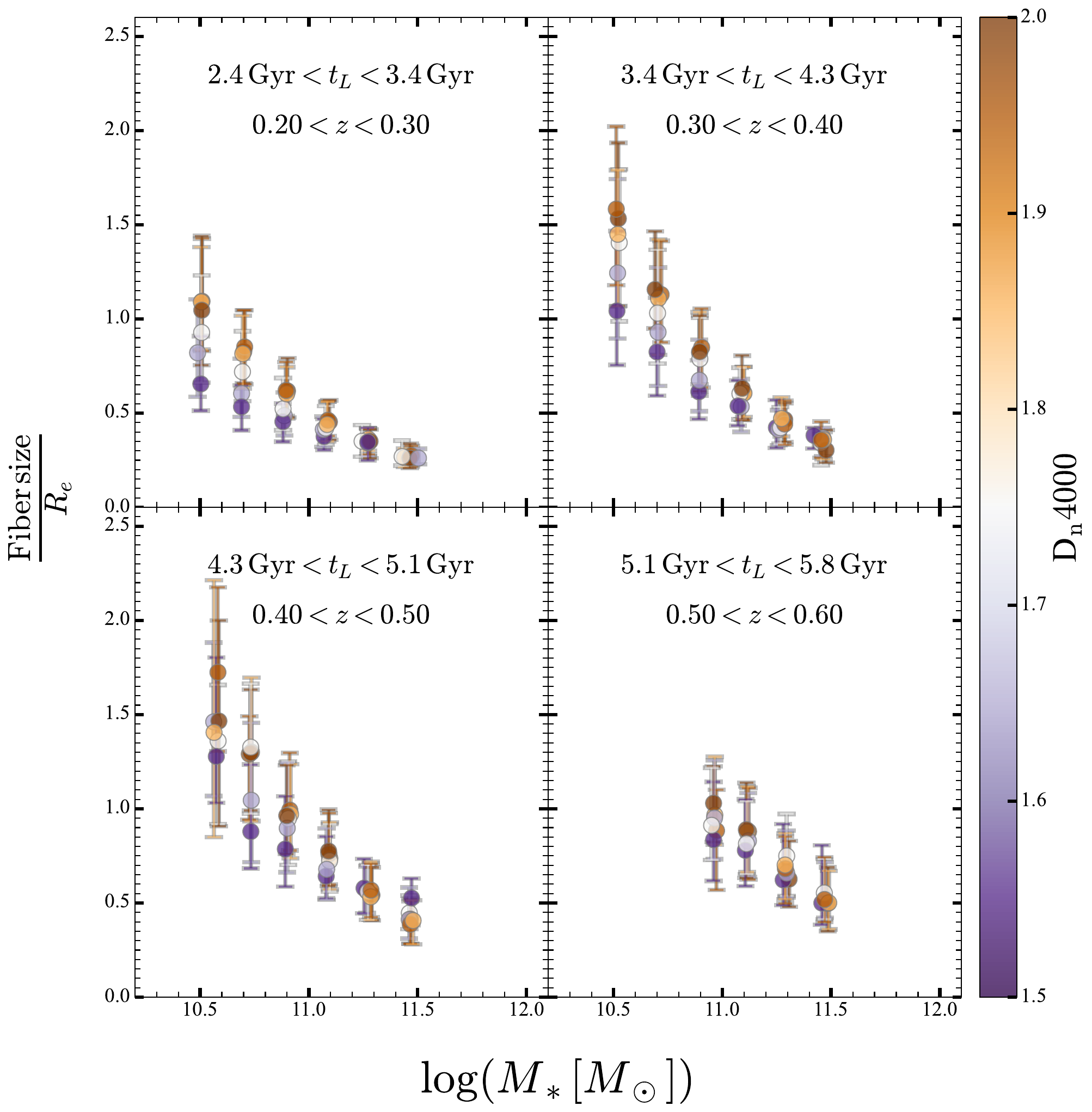}
\caption{Ratio between the Hectospec fiber aperture and the average effective radius of galaxies in the stack as a function of the central stellar mass the stellar mass - $\dn$ bins. The points are color-coded by the central $\dn$ value for each bin. The error bars indicate the (16,84) percentile range for the galaxy size distribution in the relevant bin. 
\label{f2}}
\end{figure*}

Figure~\ref{f2} shows the fiber-to-galaxy size ratio as a function of stellar mass and $\dn$ for the summed spectra in four redshift bins. As expected based on the evolution in individual quiescent galaxy sizes over this redshift range \citep[e.g.,][]{Damjanov2023}, the ratio increases with redshift at fixed stellar mass. For an average quiescent galaxy with $M_\ast\sim5-6\times10^{10}\, M_\sun$ and $\dn=1.55$ the fiber aperture coverage increases from $\sim0.4\, R_e$ at $z\sim0.25$ to $\sim0.8\, R_e$ at $z\sim0.55$. For the lowest stellar mass galaxies at $z\sim0.25$ the range in fiber coverage over the full range of $\dn$ is a factor of 2 (from $\sim0.65$ to $\sim1.3\, R_e$). For the redshift interval $0.5 - 0.6$, the range in aperture-to-size ratio with $\dn$ is small (from $\sim1\, R_e$ to $\sim1.2\, R_e$) even for the lowest mass bin. Of course, at the highest redshift the magnitude-limited HectoMAP survey includes a narrower stellar mass range.

The Hectospec fiber aperture covers at most $\sim2\, R_e$ throughout the redshift range. The average S\'{e}rsic index $n$ of the best-fit 2D models for galaxies in the mass-limited sample is $n = 4.2 \pm 1.8$. At least 60\% of the integrated light in these highly centrally concentrated profiles comes from galaxy bulges \citep[e.g.,][Appendix A, Figure A2]{Brennan2015}. Thus the light captured by the fiber aperture  ($r<2\, R_e$) is dominated by the central highest surface brightness region (the bulge); the lower surface brightness extended envelope has negligible impact because of the steep surface brightness profiles of quiescent galaxies.

\section{Full spectrum fitting of the summed spectra}\label{method}

\begin{figure}[!h]
\centering
\includegraphics[width=0.485\textwidth]{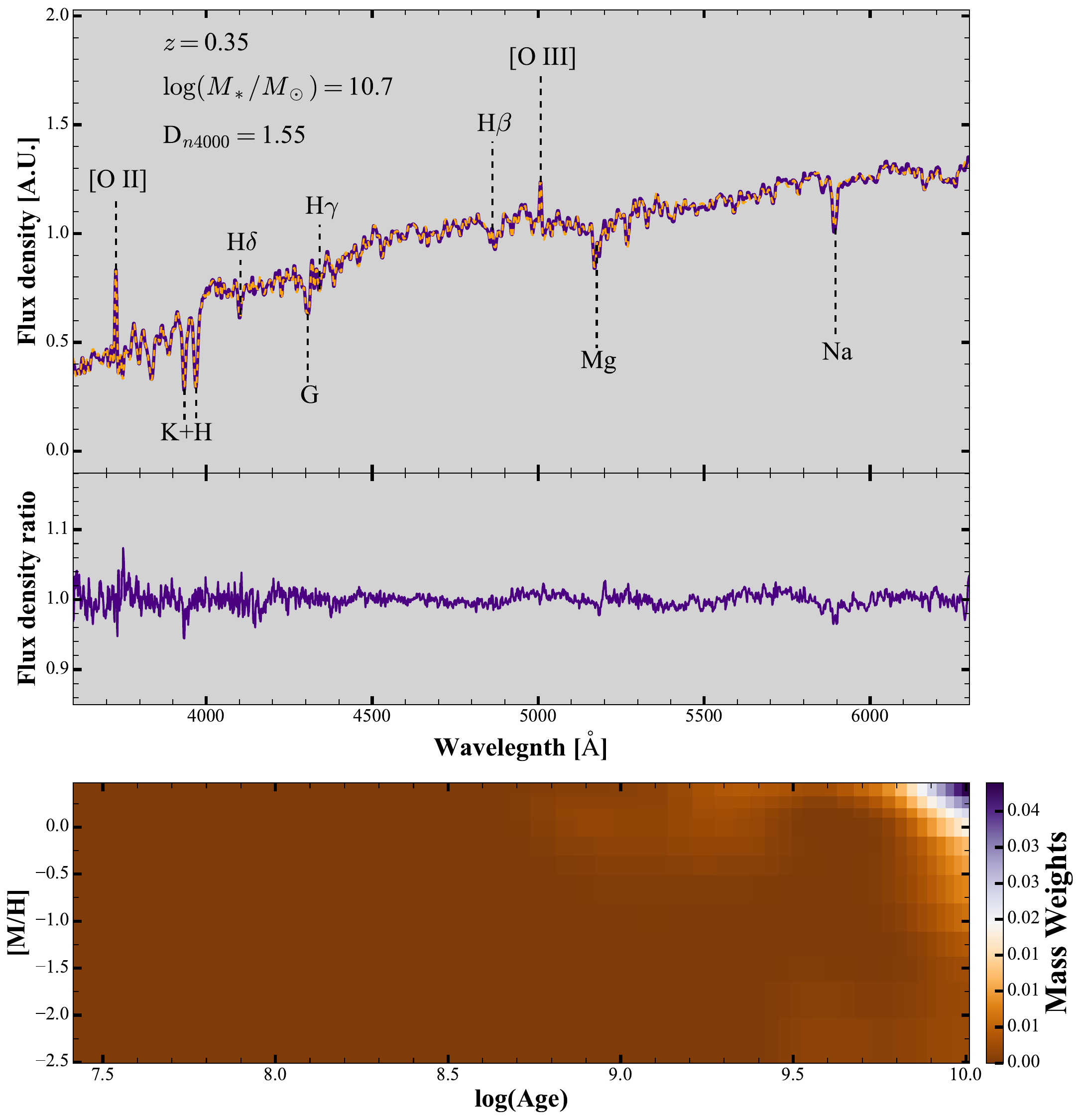}
\includegraphics[width=0.485\textwidth]{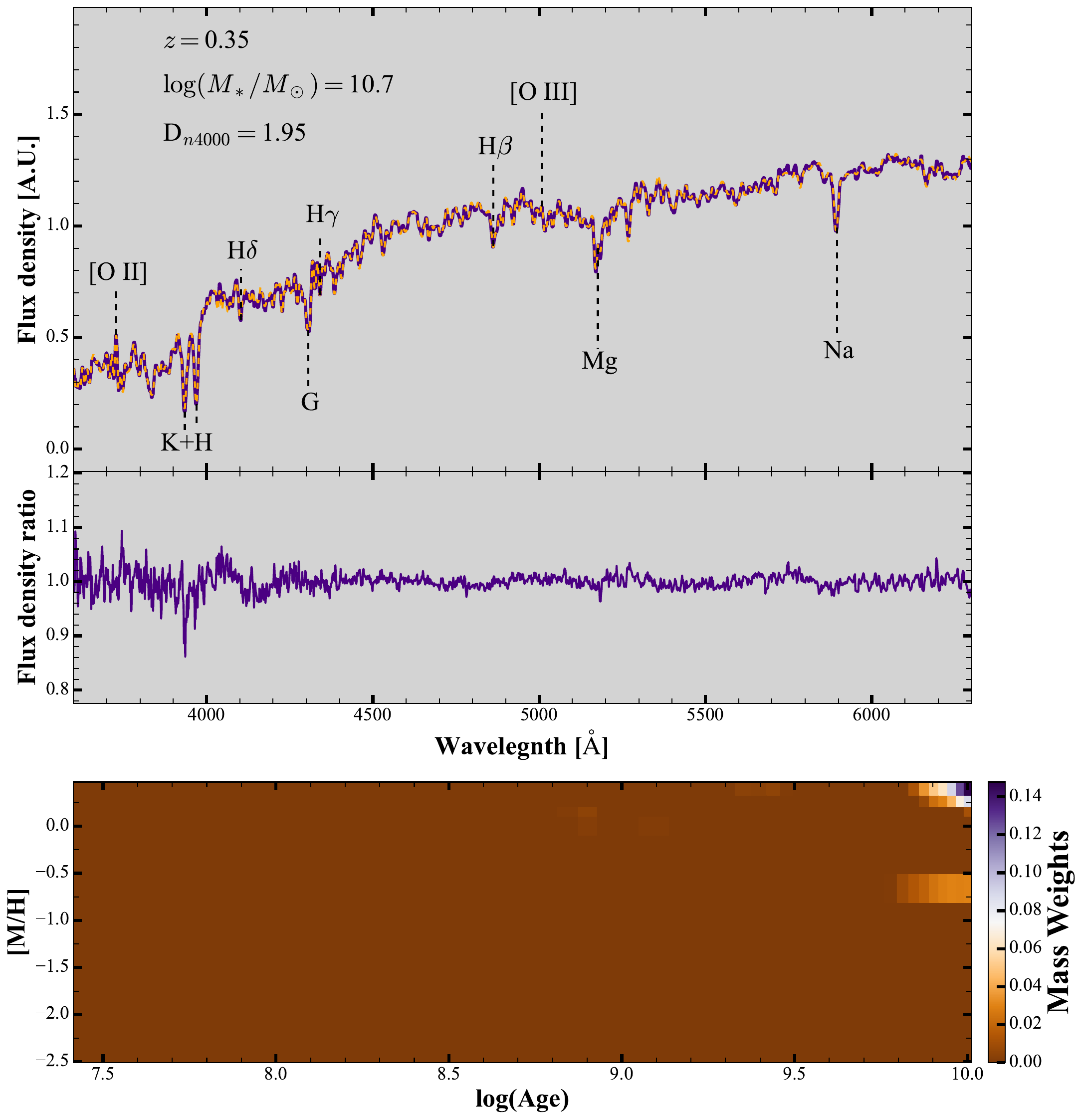}
\caption{Examples of \texttt{pPXF} fits to the average HectoMAP galaxy spectra. The legends in the top panels list the central values for the redshift, stellar mass, and $\dn$ bins for the summed spectra (Section~\ref{proc}). The top panels show the observed average spectra (solid line, violet color) and the best-fit model spectrum with maximum regularization (dashed line, orange color). The ratio between the observed and model spectra in the central panels demonstrates that the model matches the observations over the full wavelength range. The bottom panels display the mapping of weights (by light, color coding) over the stellar population age - metallicity grid of SSP models contributing to the model spectra in the top panels. \label{f3}}
\end{figure} 

We use the Penalized PiXel-Fitting (\texttt{pPXF}) software \citep[][v8.2.3]{cappellari2004,Cappellari2017,Cappellari2023} to derive average properties of the stellar and gas components from the summed HectoMAP spectra. From the highest to the lowest redshift bin in HectoMAP, the (rest-frame) wavelength coverage varies between $[2300,5300]$~\AA\ and $[2800,6500]$~\AA. The common wavelength coverage for all of the average spectra includes a number of absorption features (Ca K, Ca H, H$_\gamma$, G-band, H$_\beta$, Mg, Na) and emission lines ([OII] and [OIII], in addition to the Balmer series). 

The stellar templates we use to fit the average galaxy spectra are linear combinations of E-MILES simple stellar population models \citep[SSP,]{Vazdekis2016} based on the \citet{Salpeter1955} initial mass function (IMF; power-law with slope $\Gamma=1.3$) and BaSTI isochrones \citep{Pietrinferni2004, Vazdekis2015}. The spectral resolution of the E-MILES models in the wavelength range overlapping the rest-frame wavelength coverage of the HectoMAP spectra ($3541.4-8950.4$~\AA) is 2.51~\AA. The linear dispersion of the synthetic spectra ($\Delta\lambda=0.9$~\AA\ pixel$^{-1}$) is well matched to the wavelength pixel resolution of the HectoMAP average spectra. The $53\times12$ grid of stellar population properties for SSPs covers a [0.03,14] Gyr interval in age and a $[-2.27,0.40]$ interval in metallicity (M/H, where M/H$=0.0196$ represents the solar value). Uniform sampling, with a $\Delta$(Age)$=0.5$~Gyr step, of SSP model ages between 4 and 14~Gyr is critical for recovering the oldest average stellar population ages from the spectra of HectoMAP quiescent galaxies.

Following the recommendations in the \texttt{pPXF} documentation\footnote{\url{https://www-astro.physics.ox.ac.uk/~cappellari/software/ppxf_manual.pdf}}, we run the fitting procedure in two steps. In the first step, we do not include regularization, an addition to the procedure that provides a smoother distribution of weights in the best-fit linear combination of SSPs that is consistent with the observed galaxy spectrum \citep{Cappellari2017, Cappellari2023}. Following \citet{Spiniello2021}, we use only additive Legendre polynomials (by setting the keywords ADEGREE=10 and MDEGREE=-1) for correcting the continuum shape during fitting. This step is part of the procedure for computing galaxy velocity dispersions (Damjanov et al. in prep); all of the spectral parameters derived from the summed spectra then result from the same processing of the data.

We then rescale the input noise spectrum by $\sqrt{\chi^2/\mathrm{DOF}}$, where $\chi^2$ is the minimum value of the quantity from Eq. 18 in \citet{Cappellari2017} and $\mathrm{DOF}$ corresponds to the number of spectral pixels included in the fit. After scaling, we set the best-fit $\chi^2/\mathrm{DOF}$ (without regularization) to one.

The next step in the procedure is iterative. First, we replace the additive with multiplicative Legendre polynomials (ADEGREE=-1 and MDEGREE=10) to correct the shape of the continuum; the additive polynomials can affect the strength of the model spectral features. We then run a series of \texttt{pPXF} fits, increasing the level of regularization each time, until $\chi^2/\mathrm{DOF}$ reaches $1+\sqrt{2/\mathrm{DOF}}$ (i.e., the value one standard deviation away from the normalized mean of the $\chi^2$ distribution). As described in Section 3.5 of \citet{Cappellari2017}, the corresponding regularization value (the keyword REGUL) is the maximum that gives the smoothest distribution of weights for the SSPs in the best-fit model galaxy spectrum. A large suite of studies that include full spectrum fitting of quiescent galaxy spectra with \texttt{pPXF} vary the REGUL keyword to account for degeneracies in reproducing observed spectra with a combination of single-age single-metallicity stellar population components \citep[e.g.,][]{Wilkinson2015, Guerou2015, McDermid2015, Boardman2017, Baldwin2018, Saracco2020, Spiniello2021}.

Figure~\ref{f3} illustrates the results of full spectrum fitting for two average HectoMAP galaxy spectra from Figure~\ref{A1}. Both spectra represent galaxies in the $0.3<z<0.4$ redshift bin and the $10.8<\log(M_\ast/M_\sun)<11$ stellar mass bin. The left-hand side panels show the best fit (with maximum regularization) for the average spectrum of galaxies in the $1.5<\dn<1.6$ bin and the panels on the right show the average spectrum of galaxies with $1.9<\dn<2.0$. 

The prominent emission features (labeled, along with the absorption lines, in the top panels of Figure~\ref{f3}) are correlated with $\dn$ for galaxies  
that contribute to the average spectrum in each bin. The average spectra of low-$\dn$ $(1.5<\dn<1.7)$ 
quiescent galaxies show prominent emission lines ([O II] and H$_\alpha$). In contrast, there are no emission lines in the average spectra of high-$\dn$ galaxies (i.e., systems with $\dn \gtrsim 1.75$). 

The best-fit model spectrum (dashed orange lines in the top panels of Figure~\ref{f3}) matches the observations (solid violet lines). In both examples, the observed-to-model-spectrum ratio (central panels of Figure~\ref{f3}) is flat and nearly featureless. Over the full wavelength interval the flux density of the model spectrum remains within $\pm10\%$ of the observed spectrum. 

The maps of SSP light weights in the bottom two panels of Figure~\ref{f3} demonstrate that galaxies segregated by $\dn$ harbor stellar populations with different properties. The average spectra of low-$\dn$ quiescent systems with $\dn < 1.7$ are best modeled by linear combinations that may include up to $\sim10\%$ of young SSPs ($\leq1$~Gyr) and up to $25\%$ of $1-2$~Gyr SSPs (see the example in the bottom left panel of Figure~\ref{f3}). In contrast, the distributions of weights for SSPs in the best-fit models for high-$\dn$ galaxies ($\dn>1.9$) include no significant contribution from SSPs with stellar ages $\lesssim1$~Gyr and up to $\sim10\%$ of $<2$~Gyr old stars. The bottom right panel of Figure~\ref{f3} illustrates the fitting result that excludes any contribution from younger SSPs in the integrated light of a $\dn\sim1.95$ galaxy. The full spectrum fitting results confirm that the selection based on $\dn$ separates galaxies dominated by the light from old stellar populations from those with a non-negligible contribution from young stars. 

Throughout the text below, galaxy ages correspond to the average mass-weighted stellar population ages. Galaxy metallicities correspond to the average mass-weighted stellar metallicities. 

\section{Stellar population properties of HectoMAP quiescent galaxies}

The color selection of HectoMAP provides a mass-complete spectroscopic sample of quiescent galaxies with $0.2<z<0.6$ \citep{Damjanov2023}. The survey area provides a uniquely large sample of galaxy spectra with moderate resolution and SNR \citep[Sections~\ref{data} and~\ref{proc},][]{Sohn2021} covering this redshift range. 

The results of full spectrum fitting to the summed spectra (Section~\ref{method}) include average stellar population and emission line parameters. In Section~\ref{ages:features} we explore the correlation between the stellar population age and spectral features sensitive to the population age and/or star formation activity ($\dn$, H$_{\delta A}$, [O II] emission-line strength). In the redshift range of the HectoMAP survey there are no other mass-complete spectroscopic samples of similar size that include (average) spectra with SNR adequate for estimating stellar population age and metallicity. 

In Section~\ref{evol:am} we explore the trends in average stellar population properties (stellar age and metallicity) with galaxy stellar mass and cosmic time. To supplement the HectoMAP redshift range, we use the SDSS \citep{Gallazzi2005, CidFernandes2005, Peng2015, Trussler2020} for redshift $z < 0.2$ where the HectoMAP sample is small. At redshifts $\gtrsim 0.6$ we use the mass-complete galaxy sample with high-SNR, high-resolution spectra at $z\sim0.8$; it covers the COSMOS field, $\sim1/50$ of the area of HectoMAP \citep[LEGA-C,][]{Vanderwel2021,Cappellari2023}.

\subsection{Stellar ages and spectral indices}\label{ages:features}

\begin{figure*}
\begin{centering}
\includegraphics[width=0.95\textwidth]{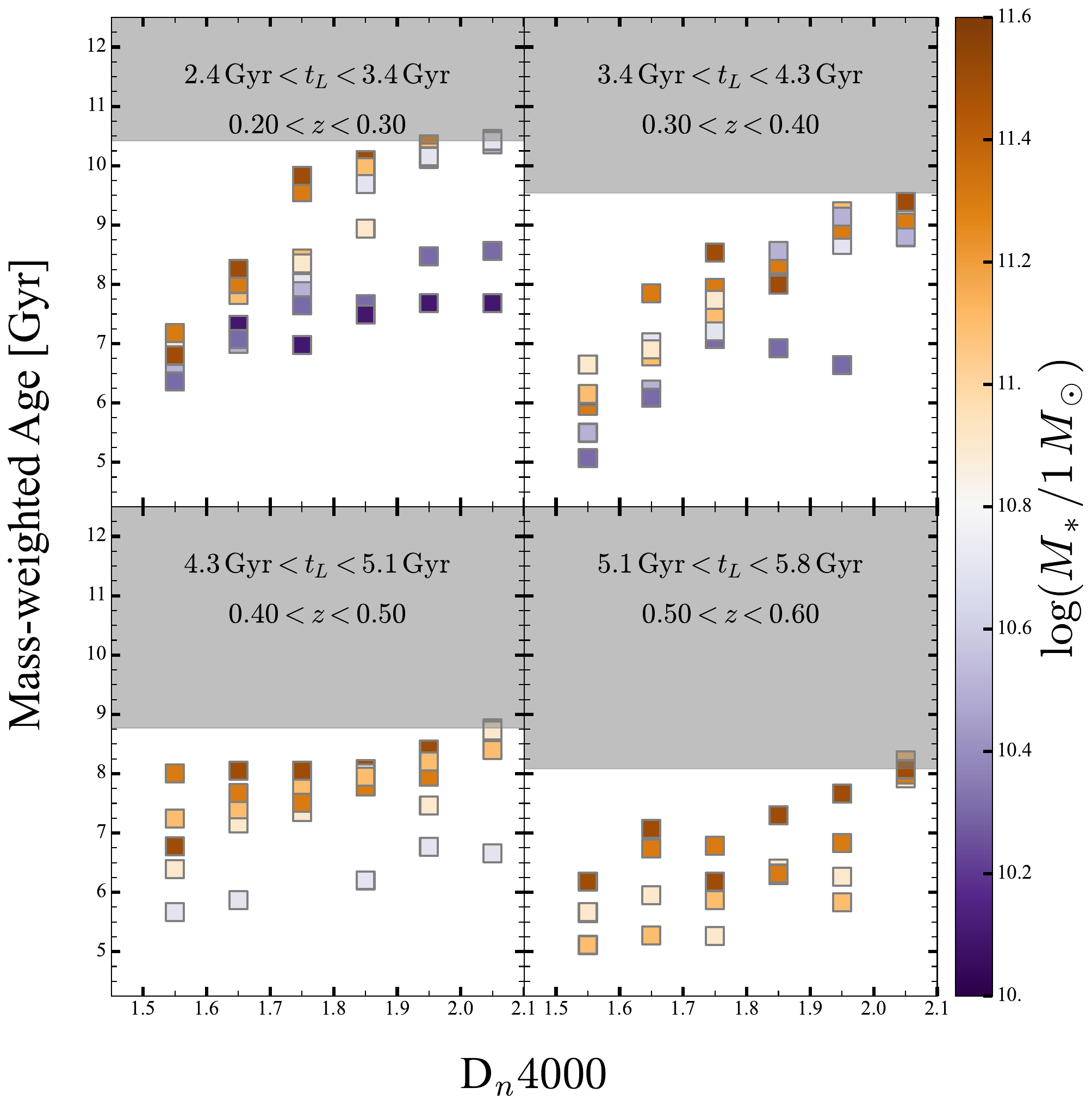}
\caption{Average mass-weighted stellar population age of the best-fit \texttt{pPXF} models vs. the central $\dn$ value of the cell for each sum (Section~\ref{proc}). The data points are color-coded by the central value of the stellar mass in each bin. Each panel shows a single redshift bin. In each panel, gray shading indicates ages exceeding the age of the universe at the maximum redshift in the interval. 
\label{f4}}
\end{centering}
\end{figure*}

The spectral indicator $\dn$ is a proxy for galaxy stellar population age. We measure  $\dn$ for individual HectoMAP galaxies (Section~\ref{data}) and use the value to group the spectra (Section~\ref{proc}).

To test the correlation between full spectrum fitting-derived galaxy stellar population age and $\dn$, we compare the average mass-weighted ages of summed galaxy spectra as a function of redshift, stellar mass, and $\dn$ values with the central $\dn$ value of the bin (Figure~\ref{f4}). Throughout the redshift interval covered by HectoMAP ($0.2<z<0.6$), there is a clear increase in the average (mass-weighted) stellar population age of the summed HectoMAP galaxy spectra with the central $\dn$ value of the relevant cells. The Spearman rank-order correlation coefficients range from $\rho = 0.53$ (for the $0.4 < z < 0.5$ redshift bin) to $\rho = 0.87$ (at $0.3 < z < 0.4$). To calculate $p-$values, we use the permutation test more appropriate for small samples than the t-distribution\footnote{https://docs.scipy.org/doc/scipy/reference/generated/scipy.stats.spearmanr.html}. In all four redshift intervals $p\leq10^{-3}$.  

Average spectra of intermediate-redshift galaxies with higher $\dn$ have older average stellar population ages based on full spectrum fitting. This relation is qualitatively equivalent to the established relation between $\dn$, the (combined) equivalent width (EW) of the H$_\delta$ and H$_\gamma$ absorption lines, and the average light-weighted stellar population age for quiescent galaxies at $z\sim0$ \citep[from the best fits to five spectral absorption features measured in SDSS DR2 spectra,][]{Gallazzi2005}. Using a spectral synthesis method similar to \texttt{pPXF}, \citet{CidFernandes2005} also find a strong correlation between the average stellar light-weighted age and $\dn$ for SDSS galaxies. The HectoMAP dataset extends the galaxy age-$\dn$ correlation to include mass-weighted ages for  $z\lesssim0.6$.  

At redshift $z\gtrsim 0.6$, galaxies samples are generally small \citep[see, e.g., Table 4 of][]{diazgarcia_2019}. A deep spectroscopic survey of several hundred massive galaxies ($M_\ast>10^{10}\, M_\sun$) at $0.6 < z < 1$ shows similar trends in mass-weighed stellar population age with $\dn$ \citep[LEGA-C survey,][]{Chauke2018}. A sample of 77 massive quiescent galaxies at $z\sim0.7$ with estimated light-weighted ages displays the same correlation \citep{Gallazzi2014}. The increase in $\dn$ that links progenitors and descendants at different redshifts \citep{Damjanov2023, Zahid2019, Damjanov2019} thus reflects the difference in the average stellar population age between younger and older quiescent galaxies at higher and lower redshift, respectively.

\begin{figure}
\begin{centering}
\includegraphics[width=0.95\textwidth]{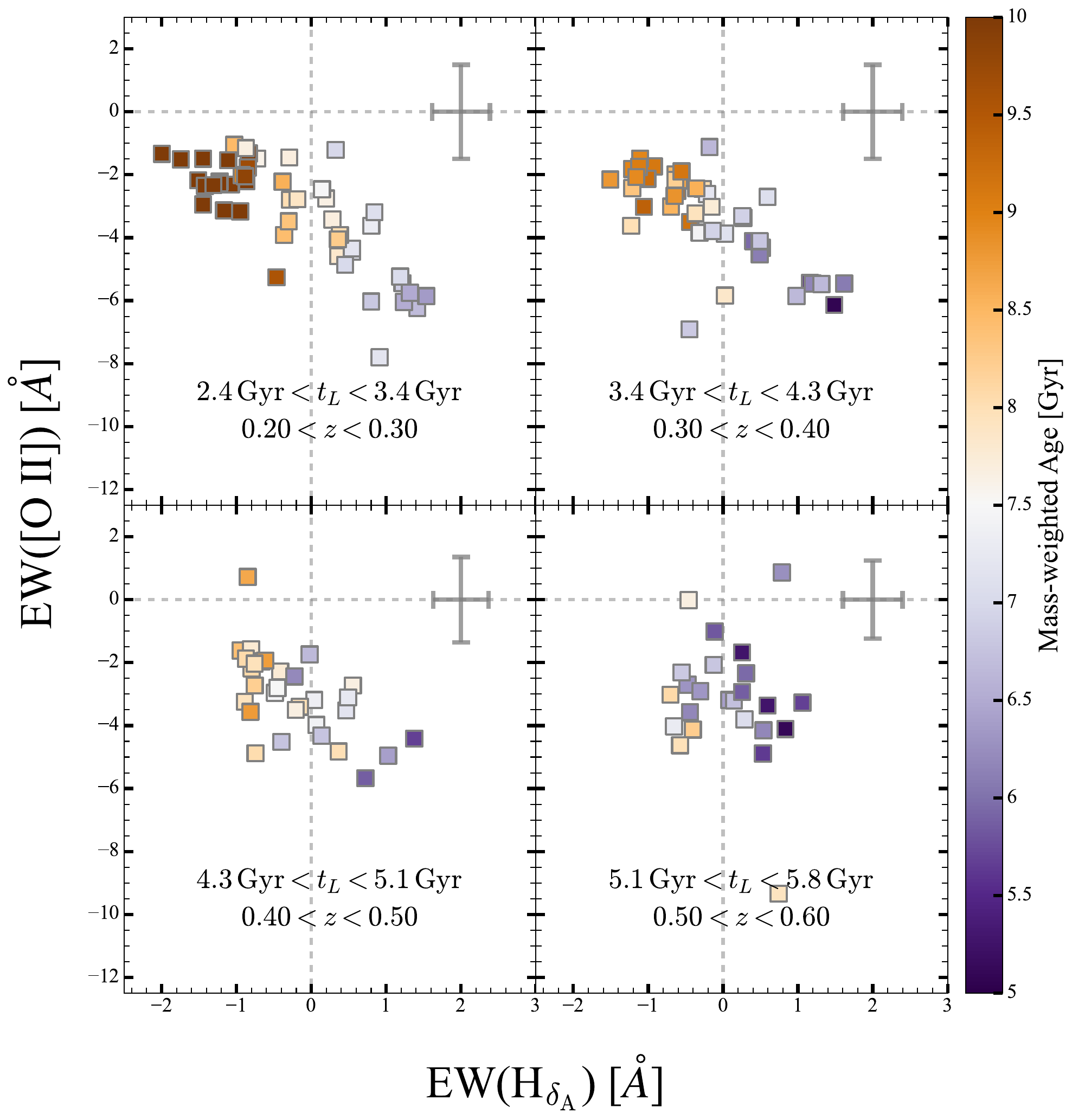}
\caption{EWs of [OII] and H$_\delta$ for quiescent HectoMAP galaxies in four redshift bins. We measure the EW directly from the average spectra of galaxies in bins of stellar mass and $\dn$. The points are color-coded by the mass-weighted average age for each spectrum. The gray bars in each panel illustrate the median errors in both EWs for galaxies in the corresponding redshift bin. 
\label{f5}}
\end{centering}
\end{figure}

The average galaxy stellar population age of quiescent galaxies correlates strongly with both $\dn$ and the strength of the Balmer absorption lines \citep[e.g.,][and their Figure 2]{Kauffmann2003}. The prominence of emission lines in the average spectra of HectoMAP galaxies with $1.5<\dn<1.6$ and their weakening with increasing $\dn$ ([O II] in all redshift bins and with H$\alpha$ in the lowest redshift bin (Figure~\ref{f1}) demonstrates non-negligible residual star formation in the central regions of young quiescent systems \citep[e.g.,][and references therein]{Zhuang2019}.

From the average spectra of quiescent HectoMAP galaxies in bins of stellar mass, $\dn$, and redshift (Section~\ref{proc}), we measure the equivalent widths (EWs) of two lines, [O II] and H$_\delta$. We use the [O II] emission line because we can trace it throughout the redshift range of HectoMAP. We select H$_\delta$ as the representative Balmer absorption line because it is isolated both from other emission and absorption lines and from strong continuum features \citep[e.g., $\dn$,][]{Goto2003}. We use the tools from Python \texttt{Specutils} package\footnote{\url{https://specutils.readthedocs.io/en/stable/fitting.html}} to perform the measurements. 

Because of the low continuum level around the [O II] line in quiescent galaxy spectra, the corresponding EW measurements are biased \citep[e.g.,][]{Yan2006}. We use simulations to estimate the minimum emission line EW that significantly exceeds the continuum EW (i.e., noise) in synthetic spectra with blue continuum levels drawn from the flux density distributions of the HectoMAP summed spectra (Appendix~\ref{EW:calib}). This procedure yields four zero points (one for each of the four redshift bins) that we subtract from all of the measured EWs. We also bootstrap the summed spectra and use the $[16, 84]$ percentile range of the resulting EW distribution to estimate errors in both line measurements. 

Figure~\ref{f5} shows the relation between [O II] and H$\delta$, color-coded by the average mass-weighted age (Section~\ref{method}). The EW of [O II] correlates strongly with the EW of H$_\delta$ throughout the redshift range of HectoMAP ($0.2<z<0.6$). The correlation coefficient is $-0.7\lesssim\rho\lesssim-0.6$ and $p\leq4\times10^{-4}$ for all but the highest redshift bin. For $0.5<z<0.6$, there is no significant correlation ($\rho=-0.17, p=0.43$) because of the narrow dynamic range in both stellar mass ($11<\log(M_\ast/M_\sun)<11.6$) and mass-weighted age ($\sim5-7$~Gyr). 

For a sample of $\lesssim1000$ rest-frame color selected quiescent galaxies at $0.6<z\lesssim1$ with high-resolution high-SNR spectra from the LEGA-C survey, \citet{Maseda2021} show that a large fraction (60\%) of individual spectra have [O II] emission above the $3\sigma$ detection threshold. They identify the [O II] line in the stacked spectrum of individual ``non-detections" as well. In a sample of quiescent galaxy spectra at $z\sim0$ \citep[GAMA survey,][]{Driver2011, Baldry2018}, \citet{Maseda2021} find the same fraction of objects with prominent [O II] (their Figure~10). HectoMAP fills the gap in EW[O II] measurements for $0.1\lesssim z<0.6$ in their figure. 

The mass-weighted average galaxy stellar population age of HectoMAP galaxies (color-coding in Figure~\ref{f5}) follows the relation between the two EWs in all four redshift bins. The full-spectrum fitting routine relies on Balmer absorption lines like H$_\delta$ to constrain the fractional contribution of single stellar populations with ages $\lesssim1-2$~Gyr. The correlation between the stellar age and the EW[O II] thus indicates that, on average, [OII] traces more recent star formation.

\subsection{Evolutionary trends in average stellar population properties}\label{evol:am}

The volume of the HectoMAP survey and the number of galaxy spectra enables, for the first time, the investigations of trends in average stellar population properties with redshift based on a mass-complete spectroscopic sample of quiescent galaxies covering the redshift range $0.2 < z < 0.6$. We use \texttt{pPXF} fitting of summed spectra with high SNR (Section~\ref{method}) to trace the mass-weighted stellar population age and metallicity as functions of stellar mass for galaxies binned by stellar mass and $\dn$ with $\dn>1.5$ and $\log(M_\ast/M_\sun)\gtrsim10.5$. We evaluate these relations in four $\Delta z=0.1$ redshift bins.

\subsubsection{Mass-weighted age}\label{evol:a}

\begin{figure}
\begin{centering}
\includegraphics[width=0.95\textwidth]{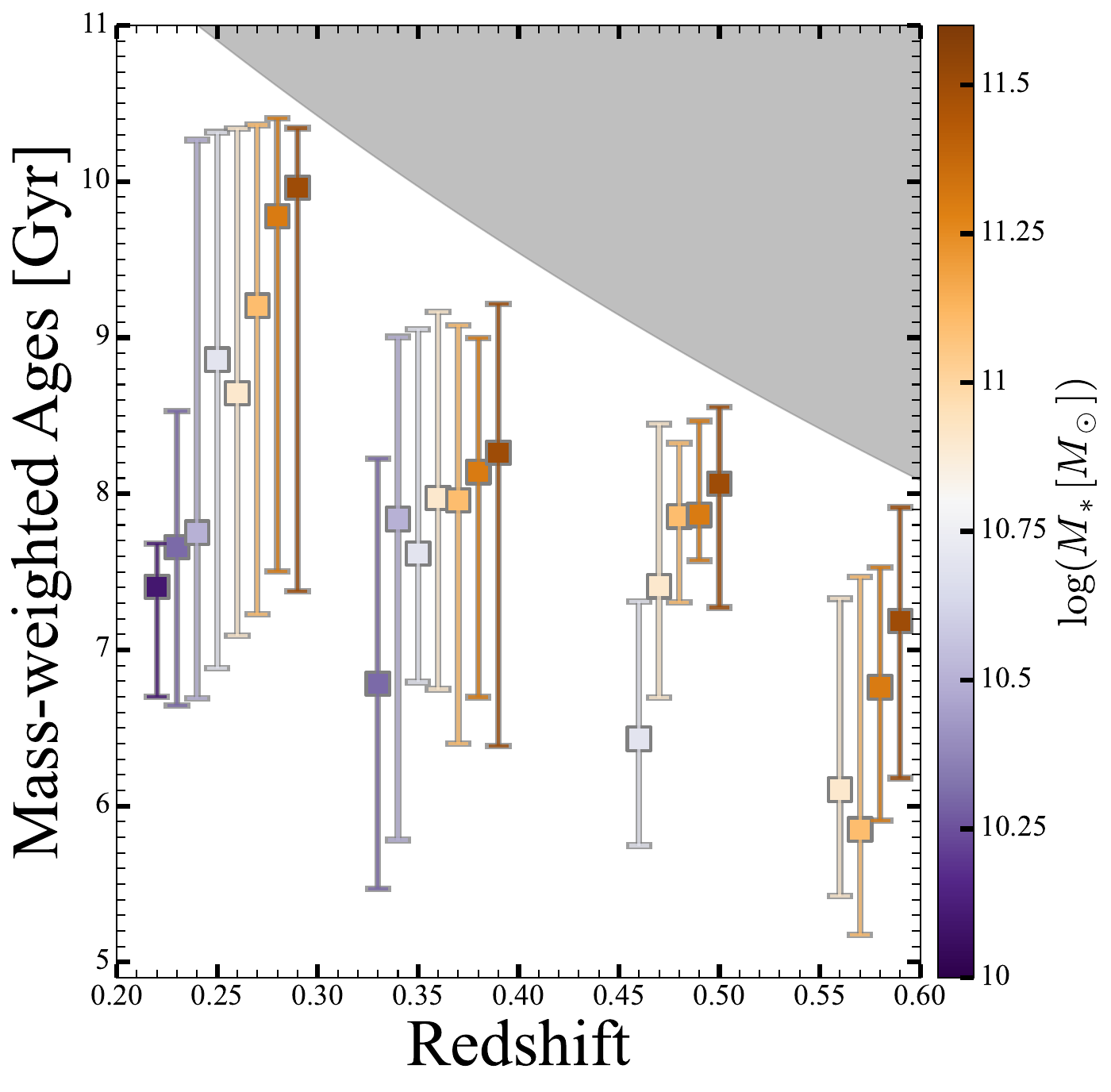}
\caption{Average mass-weighted stellar population age as a function of redshift. The symbols, indicating the median values in bins of stellar mass and redshift, are color-coded by the central value in each stellar-mass bin (Section~\ref{proc}). The four sets of points are clustered around the central values of the redshift bins (Section~\ref{proc}). The symbols show the median values in selected bins of stellar mass and redshift; the error bars show the [16, 84] percentile range. The limit of the gray region shows the age of the universe as a function of redshift. The median stellar population age of quiescent galaxies increases with increasing stellar mass. This trend extends to redshift $z\sim0.6$.
\label{f6}}
\end{centering}
\end{figure}

Figure~\ref{f6} shows mass distributions of average mass-weighted stellar population ages of quiescent systems and their redshift evolution. Grouped around the central value of the redshift bin, each set of colored squares with error bars shows the median value of the galaxy age and the corresponding [16, 84] percentile range for galaxies segregated by stellar mass. 

As expected, the average stellar population age of quiescent galaxies increases with decreasing redshift. Within individual stellar mass bins the correlation coefficient is $\rho\sim-1$ and $p\leq0.05$ for all stellar mass bins that are above the mass limit for at least three redshift intervals (i.e., $M_\ast>4\times10^{10}\, M_\sun$). For stellar masses that we trace only up to $z\sim0.4$ ($1.5\times10^{10}\, M_\sun<M_\ast<4\times10^{10}\, M_\sun$), $\rho\sim-0.8$ and $p\sim0.2$. At fixed stellar mass, all quiescent systems with stellar ages measured over the full redshift range ($0.2<z<0.6$) have galaxy ages that increase with decreasing redshift. 

Within individual redshift intervals, the median value of the average mass-weighted stellar population age increases with stellar mass at every redshift. The correlation coefficient, $\rho\gtrsim0.9$, is statically significant (with a corresponding $p-$value $\lesssim0.05$) for all redshift intervals except $0.5<z<0.6$, where $\rho\sim0.8$ and $p\sim0.3$. However, the age spread in all mass bins at fixed redshift is large and most [16, 84] percentile age intervals (error bars in Figure~\ref{f6}) overlap. Finally, in each redshift bin, the oldest average stellar population ages (corresponding to the highest $\dn$ values, Figure~\ref{f4}) are within 10\% of the age of the universe at the reference redshift.

The increase in average stellar population age with stellar mass extends from $z\sim0$ \citep[e.g.,][]{Gallazzi2005, Gallazzi2014, Peng2015, Trussler2020} to $z\sim0.6$.  Similar investigations reaching $z<1$ are based on much smaller spectroscopic samples \citep[e.g.,][]{Gallazzi2014, Choi2014}. For $z\sim0.8$ and $M_\ast>3\times10^{10}\, M_\sun$, more massive quiescent systems have later formation times \citep[the cosmic time when 50 or 90\% of stellar mass is assembled; LEGA-C,][]{Kaushal2024}. The HectoMAP results underscore the large spread in ages and the age-stellar mass relation that \citet{Kaushal2024} obtain at slightly higher redshift based on $\sim1000$ individual galaxy spectra. 

The HectoMAP survey provides a large sample $0.2<z<0.6$ that connects the redshift evolution of quiescent galaxies at $z<0.3$ from SDSS \citep[e.g.,][]{Citro2016} to future high$-z$ constraints. Existing spectroscopic samples at $z>1$ are too small to support segregation of quiescent systems by stellar mass \citep[e.g.,][]{Morishita2019}. Current estimates of quiescent galaxy ages at $z > 0.6$  suggest that the trends in the HectoMAP population extend to higher redshift \citep[e.g.,][]{Fumagalli2016, EstradaCarpenter2019, Lonoce2020, Beverage2021, Borghi2022, Tacchella2022}. Planned extensive high redshift surveys will enable determination of mass-weighted ages for subpopulations of quiescent systems that complement and extend the HectoMAP results. 

\subsubsection{Mass-weighted matallicity}\label{evol:m}

\begin{figure}
\begin{centering}
\includegraphics[width=0.95\textwidth]{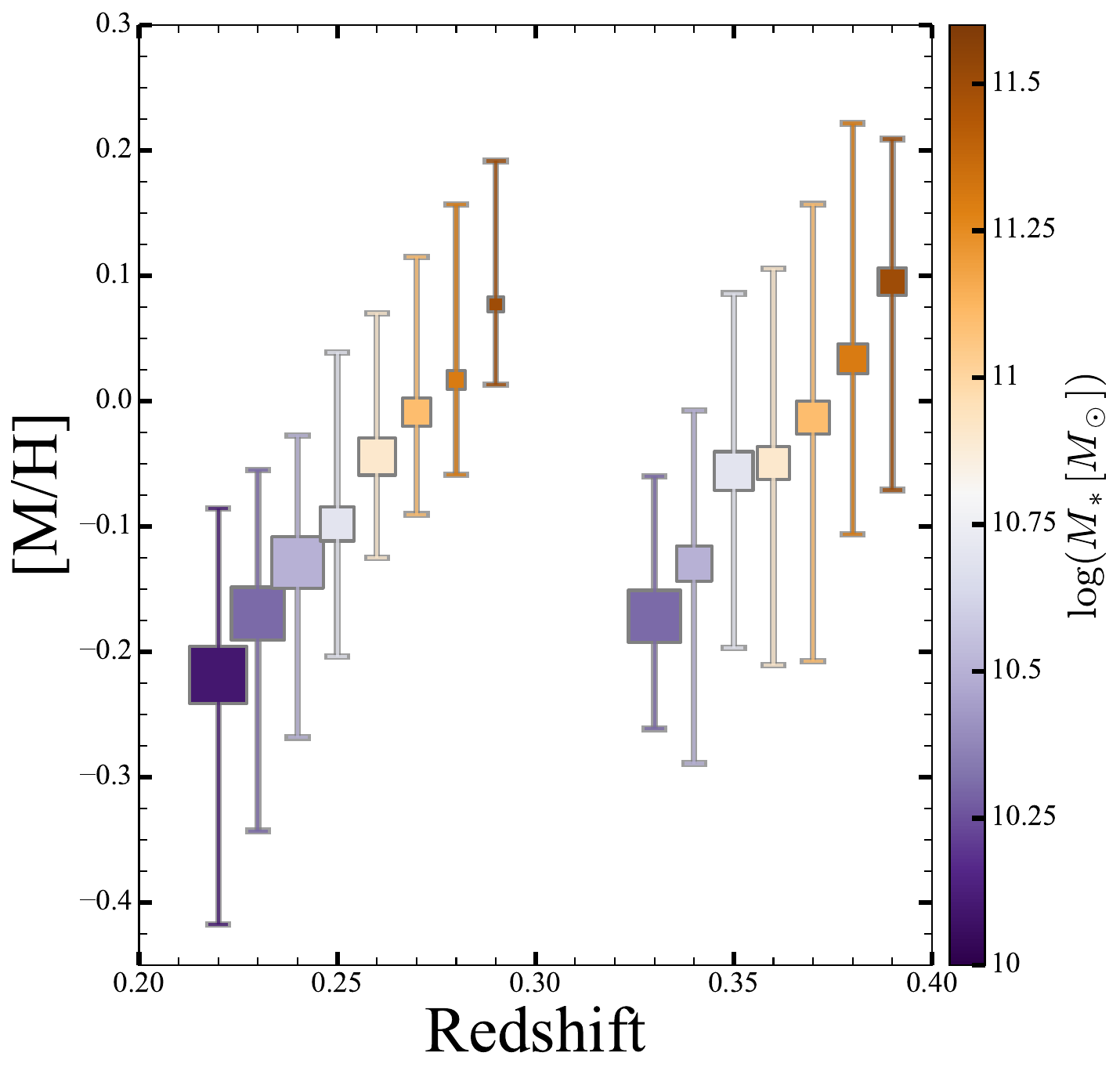}
\caption{Average mass-weighted stellar metallicity (similar to Figure~\ref{f6}). The symbol size indicates the median mass-weighed stellar population age of quiescent galaxies in each stellar mass and redshift bin. For younger systems, the symbols are larger. The increase in stellar metallicity of quiescent systems with stellar mass extends from $z\sim0$ (based on mass-complete samples from SDSS) to redshift $z\lesssim0.4$.
\label{f7}}
\end{centering}
\end{figure}

Figure~\ref{f7} shows the distribution of average mass-weighted stellar metallicity derived from the stacked spectra (Section~\ref{method}). The median mass-weighted age (symbols from Figure~\ref{f6}) determines the symbol size. Larger squares correspond to younger median ages. The small range of stellar masses HectoMAP includes at $z>0.4$ prevents constraints on trends in metallicity at higher redshift. Thus in Figure~\ref{f7} we show metallicity distributions in the only two redshift bins ($0.2<z<0.3$ and $0.3<z<0.4$) that include a wide range of stellar masses (i.e., $M_\ast>10^{10}\, M_\sun$).  

Within the two redshift bins, the median value of the average galaxy metallicity increases with stellar mass. The stellar mass - stellar metallicity trend, based on mass-complete samples covering a large range in stellar mass, thus extends from $z\sim0$ \citep[SDSS,][]{Gallazzi2005, Gallazzi2014, Trussler2020} to intermediate redshift ($z\sim0.4$). 

At $0.2<z<0.4$, we include the least-massive quiescent galaxies that are also the youngest quiescent systems in two $\Delta z=0.1$ redshift intervals (denoted by the largest squares in two clusters of symbols in Figure~\ref{f7}). These galaxies have the largest [O II] {\it and} H$_{\delta_A}$ EWs in Figure~\ref{f5}. The lowest-mass quiescent systems at $z<0.4$  ceased global star formation most recently. A non-negligible level of residual star formation in low-mass systems (with the associated feedback from young massive stars), in combination with their shallower potential compared to more massive quiescent objects,  accounts for the observed behavior \citep[e.g.,][]{Vaughan2022}. The presence of young low-mass galaxies drives the mass dependence of metallicity in quiescent systems at $0.2<z<0.4$. 

\section{Timescales for assembly of intermediate-redshift quiescent galaxies}\label{discussion}
 
\begin{figure*}
\begin{centering}
\includegraphics[width=0.975\textwidth]{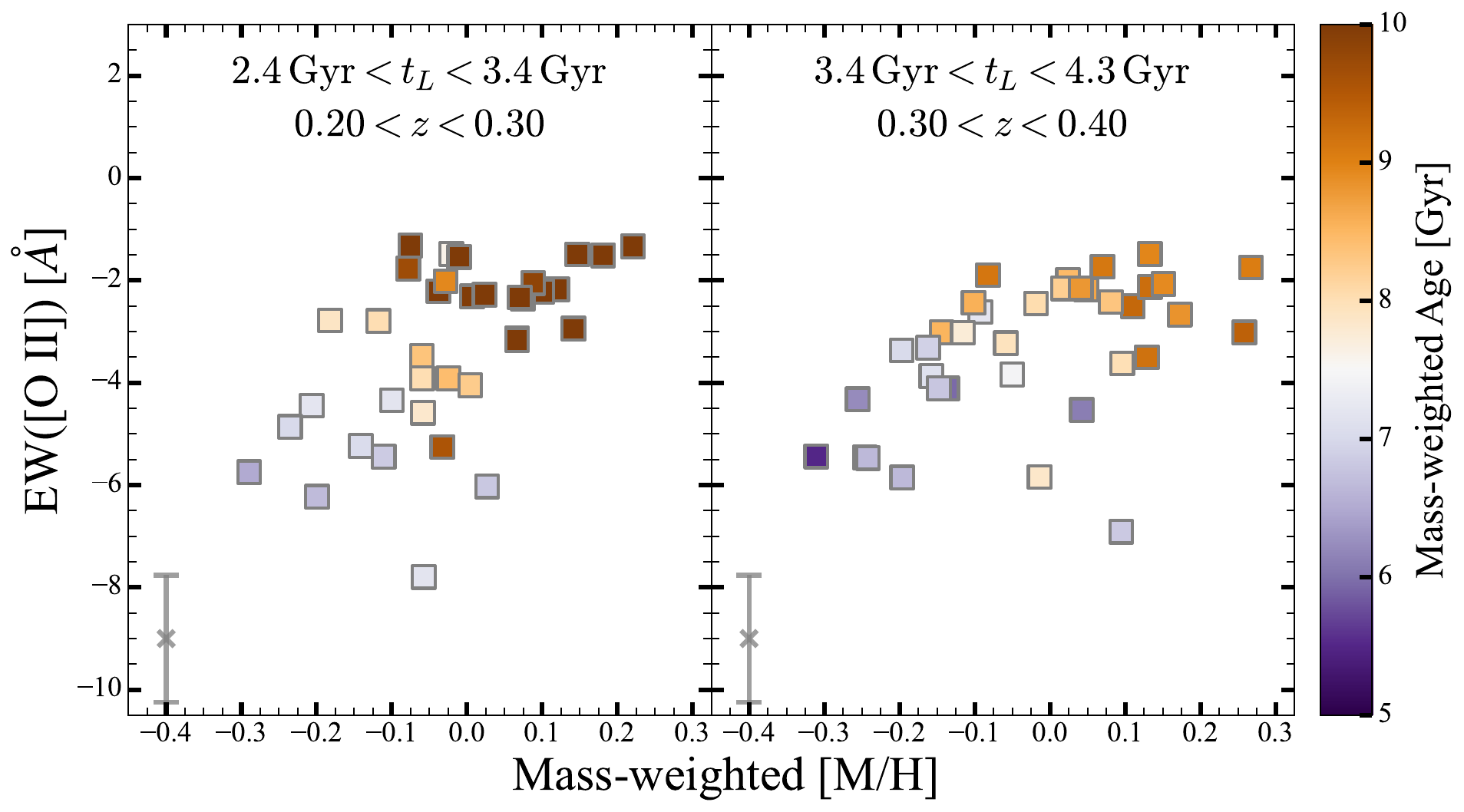}
\caption{The emission line index [O II] EW vs. average stellar metallicity [M/H] for the summed spectra of HectoMAP galaxies with $0.2<z<0.4$. These redshift intervals cover the largest range in stellar mass ($M_\ast>10^10\, M_\sun$). The symbols are color-coded by the average stellar population age (Figure~\ref{f7}). The error bar in the upper right corner of each panel shows the median error in [O II] EW within the corresponding redshift interval.
\label{f8}}
\end{centering}
\end{figure*}

In Section~\ref{ages:features}, we demonstrate tight correlations between the average stellar population ages from full spectrum fitting and the age-sensitive spectral indices measured directly from the summed spectra of quiescent ($\dn>1.5$) galaxies with $0.2<z<0.6$ (Figure~\ref{f4}). The set of (summed) high-SNR spectra confirms the use of $\dn$ as a proxy for galaxy age in individual (lower SNR) spectra and enables a connection between high-redshift progenitors and their descendants at lower redshift \citep[e.g.,][] {Hamadouche2022,Damjanov2023}. 

We further extend the correlation between inferred and measured stellar population properties to the EWs of spectral lines that are sensitive to recent star formation  ([O II] emission) and $1-2$~Gyr-old stellar populations (H$_{\delta_A}$, Figure~\ref{f5}). As in large surveys at $z\sim0$ \citep[SDSS,][]{Gallazzi2005, CidFernandes2005}, HectoMAP demonstrates a clear correlation between the strengths of emission lines and Balmer absorption lines for quiescent systems up to $z\lesssim0.6$. The mass-weighted ages follow this relation closely, with younger systems exhibiting a stronger pair of emission-absorption features. 

Recent star formation activity (traced by prominent emission lines in the visible wavelength regime) contributes to the global metal content of a galaxy inferred from full spectrum fitting. We thus explore the relation between [O II] emission and the average mass-weighted stellar metallicity in the average summed HectoMAP spectra (Figure~\ref{f8}). We restrict the discussion to the redshift interval where HectoMAP includes the largest range of stellar masses. In Section~\ref{evol:m}, we demonstrate that the stellar mass-metallicity relation from $z\sim0$ extends to $0.2<z<0.4$ and that in both $\Delta z=0.1$ bins of this interval more metal-poor quiescent galaxies harbor, on average, younger stellar populations (Figure~\ref{f7}). 

As expected based on the trends in Figures~\ref{f5} and \ref{f7}, the correlation between the [O II] EW and the inferred stellar metallicity at $z\lesssim0.4$ is strong ($\rho~\sim0.5$ with corresponding $p-$values$<0.002$, Figure~\ref{f8}). At these redshifts, the most metal-poor quiescent systems exhibit the strongest [O II] emission lines. Because of the positive correlation between the [O II] EW and average mass-weighted age, these are also the youngest quiescent systems in HectoMAP. 

The HectoMAP summed spectra paint a coherent picture of evolution in the stellar population properties of quiescent galaxies at intermediate redshift. The average mass-weighted age in the central regions of quiescent systems ($0.5R_e-2R_e$, Figure~\ref{f2}) increases monotonically with $\dn$ (Figure~\ref{f4}). In every $\Delta z=0.1$ redshift bin, the summed spectra of galaxies with the highest $\dn$ have average stellar population ages close to the age of the universe at the central redshift of their bin (grey-shaded regions in Figure~\ref{f4}). Quiescent galaxies at intermediate redshift thus exhibit a range of mass assembly histories; those with $\dn\sim2$ form the bulk of their stellar mass early in cosmic history.

The average stellar population age increases steadily with cosmic time for galaxies of all stellar masses (Figure~\ref{f6}). We observe the aging of stellar populations in the central regions of quiescent systems from $z\sim0.6$ to $z\sim0.2$. Within individual redshift bins there is a weak increase in the median value of the mass-weighted age distribution as a function of stellar mass (square symbols in Figure~\ref{f6}). Quiescent systems with stellar masses $M_\ast>5\times10^{10}\, M_\sun$ show no significant trend in age with stellar mass. The absence of a trend is consistent with the evolutionary scenario where massive galaxies form a large fraction of their stellar mass earlier than their lower-mass counterparts \citep[downsizing in mass; e.g.,][]{Thomas2010}. 

In the redshift interval $0.2<z<0.3$ (where the range of stellar masses extends to $\sim\times10^{10}\, M_\sun$), downsizing is visible in the difference between the average mass-weighted age distribution for quiescent galaxies below and above $M_\ast\sim3\times10^{10}\, M_\sun$. The lowest-mass galaxies  ($M_\ast\sim1-2\times10^{10}\, M_\sun$) have a narrower range of average mass-weighted ages and the lowest central (median) value in their age distribution. This behavior results from recent star formation episodes (as traced by more prominent [O II] emission, Figure~\ref{f5}). In contrast, all of the quiescent systems with $M_\ast\gtrsim3\times10^{10}\, M_\sun$ at this redshift exhibit a similarly wide range of mass assembly histories and older median stellar population ages. 

The stellar mass trends in the distribution of average mass-weighted metallicity (Figure~\ref{f7}) illustrate that the stellar mass-metallicity relation extends at least from $z\sim0$ to $z\sim0.4$.  Both median metallicity and its [16, 84] percentile range increase with stellar mass for $M_\ast\lesssim5\times10^{10}\, M_\sun$ galaxies (the two clusters of points on the left-hand side of Figure~\ref{f7}). Galaxies with the lowest stellar mass ($M_\ast\sim2\times10^{10}\, M_\sun$) have the lowest (subsolar) metallicites. The correspondingly strong [O II] emission lines and young average stellar population ages of these low-mass metal-poor systems signal more recent star formation activity in their central regions (Figure~\ref{f8}). A combination of stellar feedback from massive young stars and the shallow central potential can explain the trend towards lower metallicites with decreasing stellar mass of quiescent systems with $M_\ast\lesssim5\times10^{10}\, M_\sun$ \citep{Looser2024}. Thus the increase in stellar metallicity with stellar mass (and average age) again underscores the more extended mass assembly history for intermediate-redshift quiescent galaxies with lower stellar mass.

\section{Conclusion}\label{conclusions}

Large redshift surveys enable the construction of representative average (summed) spectra of galaxies segregated by spectral properties derived from individual spectra (i.e., redshift, $\dn$, and stellar mass). The order of magnitude increase in the SNR of the summed spectra enables full-spectrum fitting and the extraction of average stellar population properties based on the best-fit linear combination of SSP models. 

We obtain mass-weighted stellar population ages and metallicites from 144 summed spectra representing more than 35,000 quiescent galaxies with $M_\ast>10^{10}\, M_\sun$ in the HectoMAP survey \citep{Sohn2021, Sohn2023}. We explore trends in galaxy stellar population properties with stellar mass and spectral indices in the HectoMAP redshift range, $0.2 < z < 0.6$, that uniquely fills a gap in previous studies \citep[e.g.,][]{Costantin2019, Maseda2021}.

The average mass-weighted galaxy age increases steadily with increasing $\dn$  (Figure~\ref{f4}). The $\dn$ spectral index is based on the average (or median) continuum flux in a narrow wavelength range around 4000~\AA\ (Section~\ref{data}). In contrast with full-spectrum fitting, $\dn$ can be extracted from individual spectra with modest spectral resolution and wavelenght coverage \citep[e.g., $R\sim500$ and 5550–-9650~\AA\ wavelength coverage of zCOSMOS provides $\dn$ measurements for $0.4<z<1$ redshift interval;][]{Cucciati2010}. The $\dn$ index ranks quiescent galaxies based on their relative ages (Section~\ref{discussion}) thus providing the avenue for linking progenitors and descendants over a broad redshift range (e.g., by combining different surveys).

In addition to $\dn$, average stellar population age strongly correlates with the strengths of the [O II] emission line and the H$_\delta$ absorption line measure from summed spectra (Figure~\ref{f5}). Galaxies with younger dominant stellar populations exhibit larger (absolute) EWs for both the nebular emission line ([O II]) and the absorption feature that are most prominent in the atmospheres of A-type stars (H$_\delta$). This correlation implies that the nebular emission line traced throughout the redshift interval of the survey - [O II] - is an indicator of the most recent star formation episodes in quiescent galaxies.

The strength of the [O II] emission line and the average stellar metallicity of HectoMAP galaxies at $0.2<z<0.4$ are anti-correlated (Figure~\ref{f8}). In this redshift interval, our mass-complete sample covers the largest range of stellar masses and thus provides the most extended baseline for average metallicity estimates. Together with the correlation between EW [O II] and average stellar population age, stronger EW[O II] in lower-metallicity galaxies indicates a relation between (the lack of) metal enrichment and the prominence of the most recent star formation activity in quiescent galaxies.

In all four $\Delta z=0.1$ redshift intervals and for all stellar masses, the distribution of galaxy ages is broad (Figures~\ref{f4} and~\ref{f6}). Only a fraction of the most massive galaxies in each redshift bin ($M_\ast>10^{11}\, M_\sun$) have ages similar to the age of the universe at that redshift. The median age increases with stellar mass from $z\sim0.25$ to $\sim0.55$, albeit with a large scatter (quantified by the [16,84] percentile interval). The increase in median age with stellar mass and the broad distributions of ages in the majority of stellar mass bins demonstrate a range of star formation histories for $M_\ast\gtrsim10^{10}\, M_\sun$ quiescent systems and, on average, earlier mass assembly of more massive galaxies. At all redshifts, the average ages of the lowest-mass galaxies are most affected by recent star formation episodes (Figures~\ref{f5} and~\ref{f6}).

At two redshift bins within the range $0.2<z<0.4$, distributions of metallicities in bins of stellar mass, the median metallicity increases with the central stellar mass  (Figure~\ref{f7}). Lower-mass quiescent galaxies, with (on average) larger mass fraction of younger stellar populations, have lower metallicities. Taken together with the prominence of the [O II] emission line in the spectra of younger and more metal-poor galaxies, these trends suggest that outflows driven by stellar feedback (stellar radiation, stellar winds, or supernova explosions) from most recent star formation episodes affect the metal content of lower-mass galaxies with shallower potential wells. Large samples of star-forming (progenitor) galaxies at higher redshift are critical for detailed tests of this scenario \citep{Trussler2020}.  

Spectroscopic samples at intermediate redshift trace galaxy evolution over nearly half of cosmic history. Until recently, there were no sufficiently large surveys with adequate spectral quality to explore the redshift range between  $z\sim0.2$ and $z\sim0.7$ \citep{Costantin2019}. The HectoMAP survey provides a mass-complete sample of quiescent galaxies with $0.2<z<0.6$, complementing studies at $z\sim0$. Ongoing and upcoming spectroscopic surveys \citep[DESI, PFS, WEAVE;][]{Schlafly2023,Greene2022,Iovino2023} will provide similarly large samples of both star-forming and quiescent systems at similar and higher redshift. 

\begin{acknowledgments}
 I.D. acknowledges the support of the Canada Research Chair Program and the Natural Sciences and Engineering Research Council of Canada (NSERC, funding
reference numbers RGPIN-2018-05425 and RGPIN-2024-06874). M.J.G. is supported by the Smithsonian Institution. J.S. is supported by the National Research Foundation (NRF) of Korea grant funded by the Korean Ministry of Science and Information and Communication Technology (MSIT; RS-2023-00210597). This work was also supported by the Global-LAMP Program of the National Research Foundation of Korea (NRF) grant funded by the Ministry of Education (No. RS-2023-00301976). In addition, the engagement of I.D. and J.S. in this research project has been supported by the Brain Pool Program funded by the MSIT through the NRF of Korea (RS-2023-00222051).
\end{acknowledgments}

\vspace{5mm}
\facilities{MMT Hectospec}

\software{Astropy \citep{Astropy13,Astropy18, Astropy22}, Specutils \citep{nicholas_earl_2024}, SciPy \citep{2020SciPy-NMeth}, NumPy \citep{ harris2020array}, Matplotlib \citep{Hunter:2007}, pPXF \citep{cappellari2004,Cappellari2017,Cappellari2023}}

\appendix

\restartappendixnumbering

\section{Summed spectra of HectoMAP galaxies at $0.3<z<0.6$}\label{Spectra:other}

\begin{figure*}[!h]
\centering
\includegraphics[width=0.8\textheight]{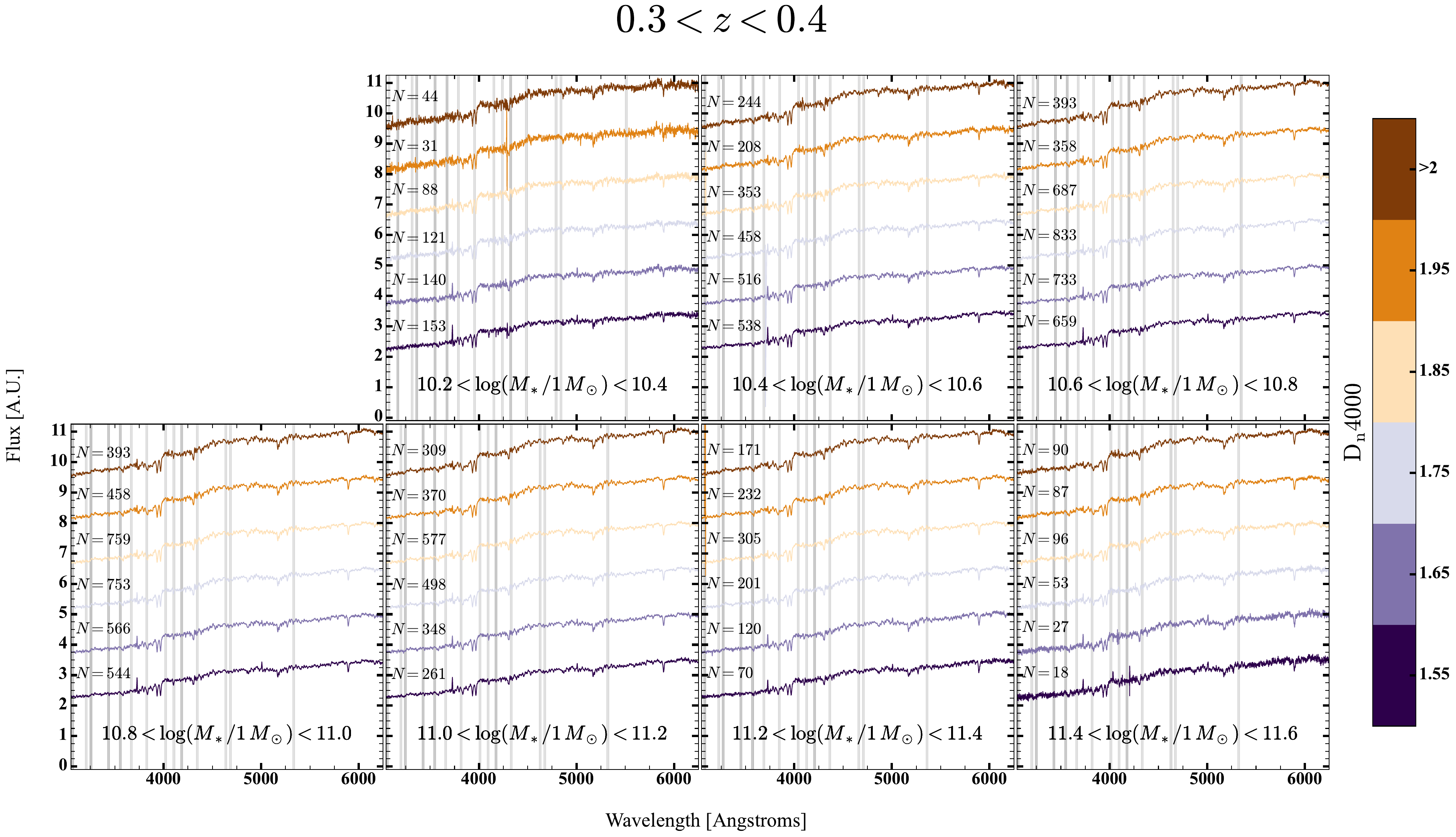}
\caption{Averaged summed spectra of HectoMAP quiescent galaxies with $0.3<z<0.4$ segregated by stellar mass (panels) and $\dn$ (colors). We include mass bins above the stellar mass limit of the survey. The number next to each spectrum indicates the total number of individual HectoMAP spectra in the stack. Gray lines show the positions of emission lines in the Mount Hopkins night sky spectrum (Massey \& Foltz 2000) in the rest frame of the $0.3<z<0.4$ stack. The average flux density of the co-adds is in arbitrary units. 
\label{A1}}
\end{figure*} 

\begin{figure*}[!h]
\centering
\includegraphics[width=0.8\textheight]{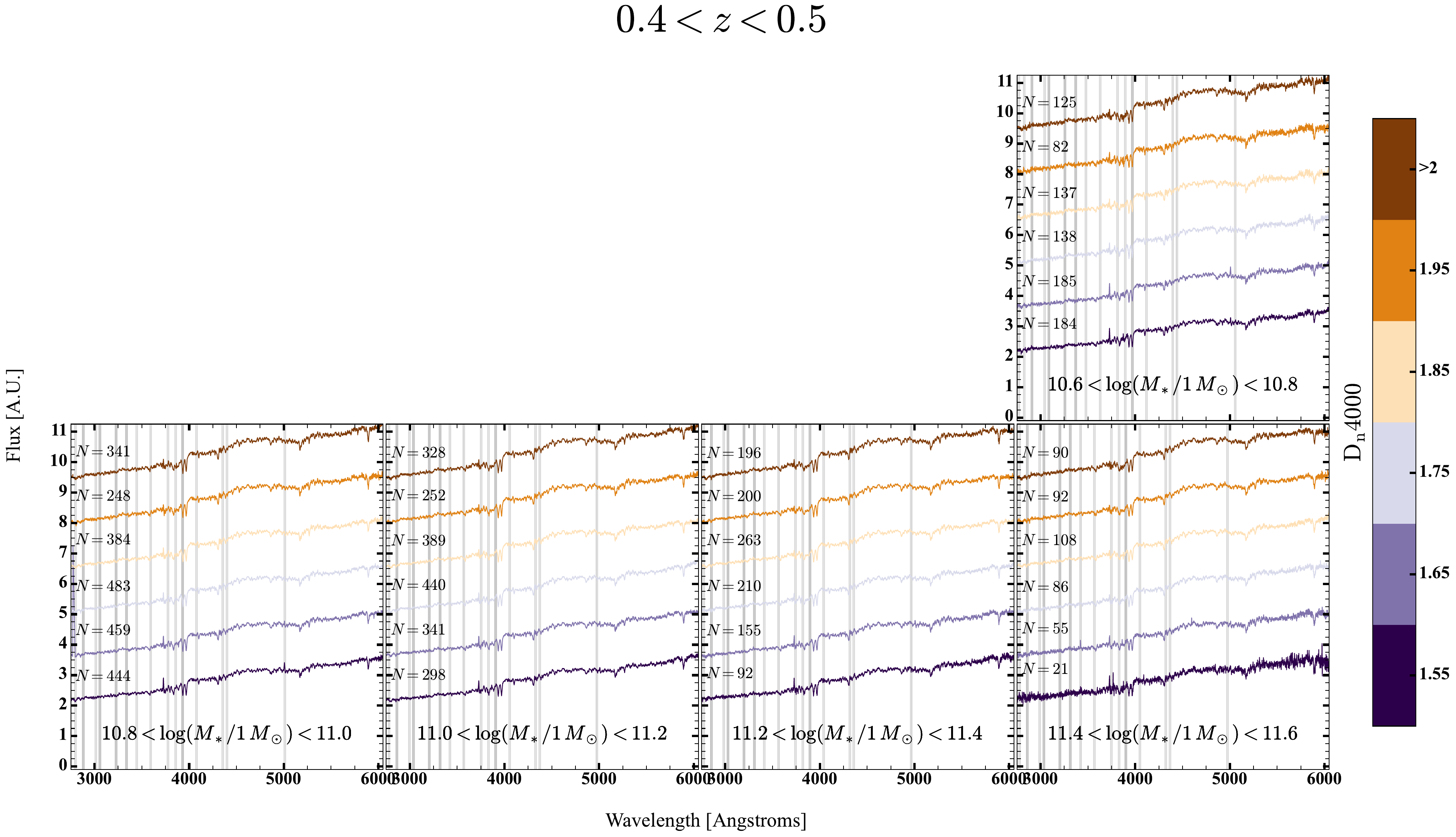}
\caption{Equivalent to Figure~\ref{A1} for $0.4<z0.5$ quiescent galaxies. We exclude the narrow redshift slice between $z=0.4$ and $z=0.42$ where the most intense sky line, [O I]$\lambda5577$, falls in the range of redshifted CaII H+K absorption features.
\label{A2}}
\end{figure*} 

\begin{figure*}[!h]
\centering
\includegraphics[width=0.8\textheight]{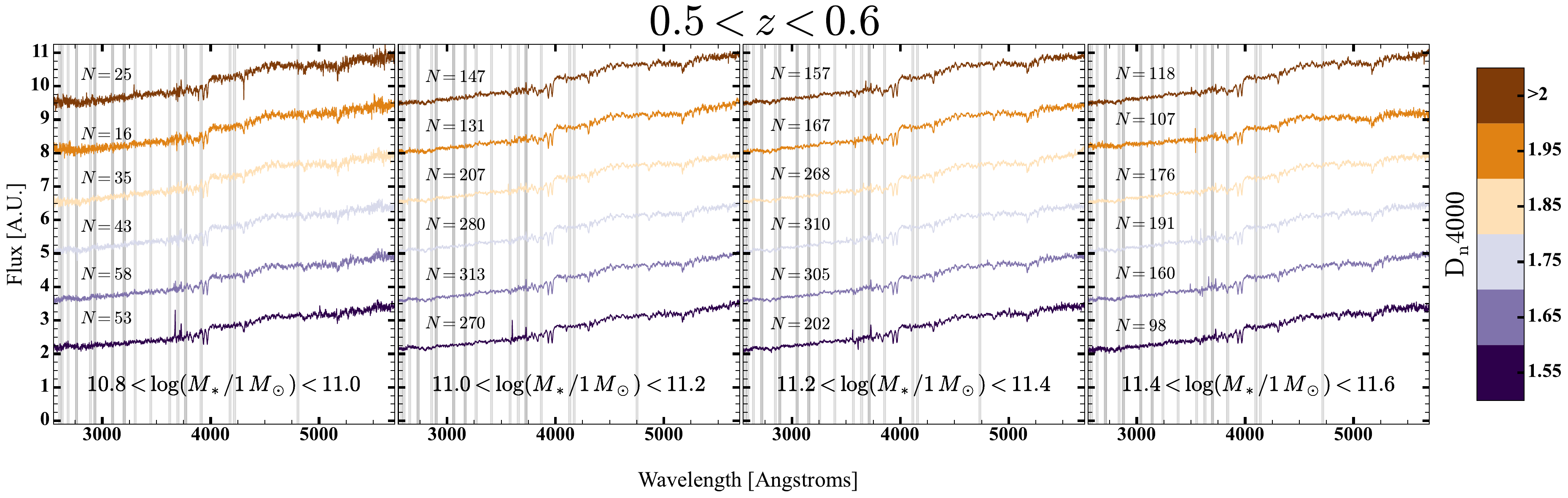}
\caption{Equivalent to Figure~\ref{A1} for $0.5<z<0.6$ quiescent galaxies.
\label{A3}}
\end{figure*} 

Section ~\ref{proc} describes the procedure for constructing high-SNR summed spectra of quiescent galaxies segregated by redshift, stellar mass, and $\dn$. Figure~\ref{f1} shows the resulting spectra for galaxies covering the redshift interval $0.2<z<0.3$. Here we display all of the additional summed spectra we use in the analysis (Figures~\ref{A1}--\ref{A3}).

\section{Calibration of the EW  of the [O II] emission line}\label{EW:calib}

\begin{figure*}
\begin{centering}
\includegraphics[width=0.95\textwidth]{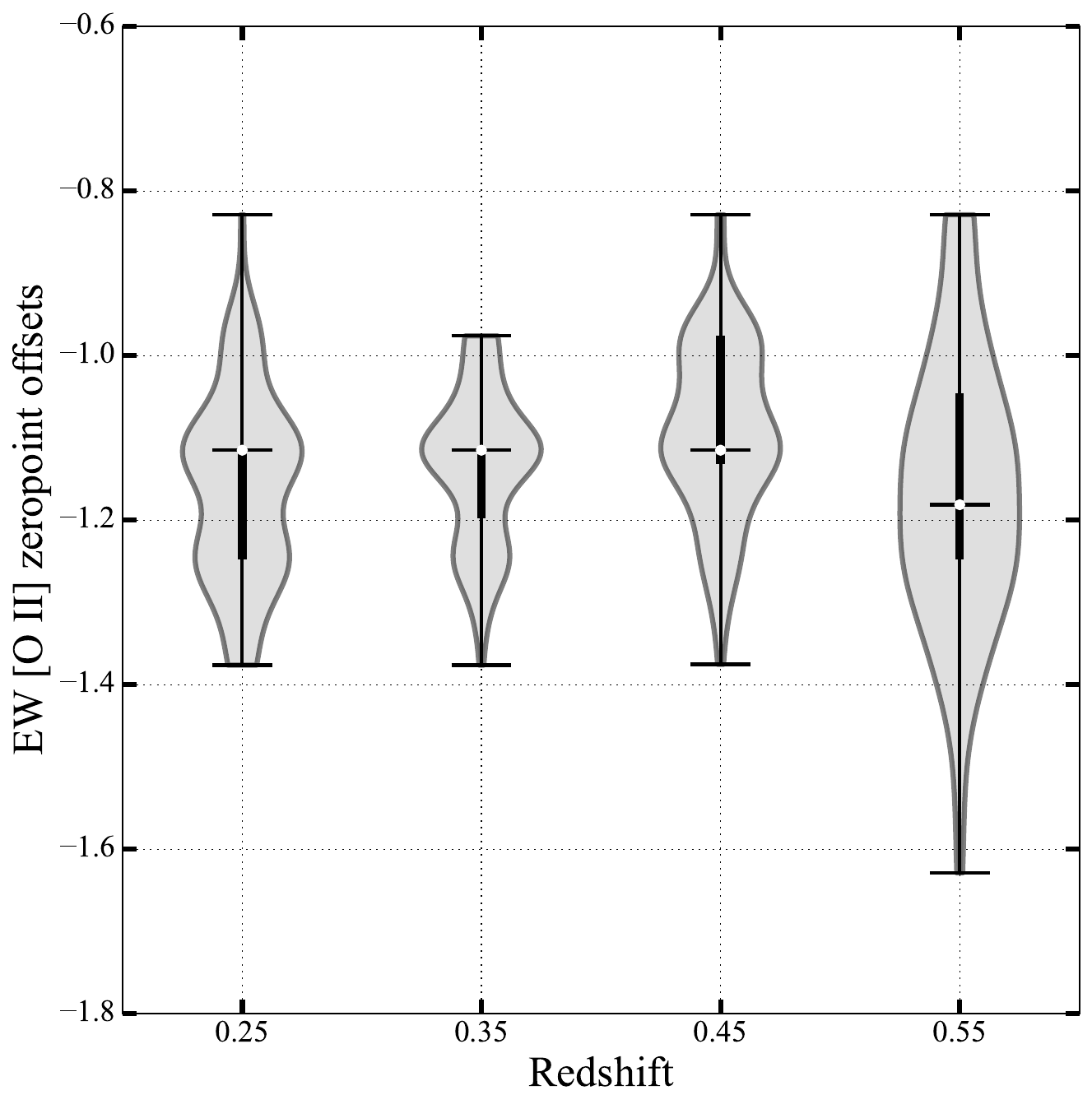}
\caption{Distribution of zeropoint offsets for [O II] EWsin four redshift bins. White circles indicate the median values and black thick lines show interquartile ranges. See text for details.
\label{B1}}
\end{centering}
\end{figure*}

We use simulations  to test and calibrate the emission-line EWs in Section~\ref{ages:features}.  Figure~\ref{f5} of  Section~\ref{ages:features}  shows the correlation between the measured EWs for [O II] emission and H$_\delta$ absorption lines for the quiescent HectoMAP sample. For the [O II]$\lambda$3727\AA\ line, the surrounding (blue) continuum is low. Large fractional Flux variations in the narrow continuum ranges on the blue and red side of [O II] line can affect the continuum estimate significantly thus producing a factor of several in the difference between the measured and true EWs. 

To investigate the effect of noise on the [O II] EW, we construct a series of simulated spectra covering the wavelength range $[3696, 3758]$\AA. This wavelength range includes the [O II] line and the continuum around it (Fisher et al. 1998).  

For each summed spectrum we create a simulated continuum by sampling from a Gaussian with a mean equal to the mean value of the flux in the continuum around the [O II] line and the standard deviation equal to the root mean square deviation (RMSD) of the continuum in that spectral range. One model at a time,  we add a series of model emission lines as Gaussian functions with a range of parameters (amplitude, mean, and standard deviation) to the simulated continuum. In each step we measure the EW of the model line both with the perfectly flat (intrinsic) and the noisy (simulated) continuum and compare the two values. The EW measured from a noisy spectrum always exceeds its intrinsic value. The intrinsic EW where the relative difference falls below 20\% is the maximum EW that may appear in a noisy continuum when there actually no real emission line. We thus subtract this zeropoint from each measured EW.

We repeat this procedure 100 times for each average spectrum; in every iteration we record the median offset for all spectra in a single redshift bin. The violin plots of Figure~\ref{B1} illustrate the distribution of median offsets for all four $\Delta z=0.1$ redshift intervals covered by HectoMAP. We select the median of the distribution (the white circle in each violin plot) as the EW measurement zeropoint for that redshift bin. The zeropoints remain very stable ($\sim-1.1$~\AA) over the full redshift range.  
%========================================
% bibliography
%========================================

%\bibliographystyle{aasjournal}
\bibliography{Summed_spectra}

\end{document}